\documentclass[twocolumn,epjc3]{svjour3}
\usepackage{graphicx}
\usepackage[numbers,sort,compress]{natbib}
\usepackage[breaklinks=true]{hyperref}
\hypersetup{
    colorlinks=true,
    linkcolor=blue,
    citecolor=blue,
    filecolor=magenta,      
    urlcolor=cyan,
    pdftitle={},
    }
\journalname{Eur. Phys. J. A}
\begin{document}
%
\title{Advancements of $\gamma$-ray spectroscopy 
of isotopically identified fission fragments with AGATA and VAMOS++}
\author{
A.~Lemasson\thanksref{addr1, e1} 
\and 
J.~Dudouet\thanksref{addr2}
\and
M.~Rejmund\thanksref{addr1}
\and  
J.~Ljungvall\thanksref{addr3} 
\and 
A.~G\"{o}rgen\thanksref{addr4} 
\and 
W.~Korten\thanksref{addr5}
} 
\thankstext{e1}{e-mail: antoine.lemasson@ganil.fr}

\institute{GANIL, CEA/DRF-CNRS/IN2P3, Bd Henri Becquerel, BP 55027, F-14076 Caen Cedex 5, France \label{addr1}
\and
Université de Lyon 1, CNRS/IN2P3, UMR5822, IP2I, F-69622 Villeurbanne Cedex, France \label{addr2}
\and 
IJCLab, Universit\'e Paris-Saclay, CNRS/IN2P3, F-91405 Orsay, France \label{addr3}
\and
Department of Physics, University of Oslo, PO Box 1048 Blindern, N-0316 Oslo, Norway \label{addr4}
\and
IRFU, CEA, Université Paris-Saclay, 91191, Gif-sur-Yvette, France  \label{addr5}
}
%
\authorrunning{"A. Lemasson et al"}
\titlerunning{"Advancements of $\gamma$-ray spectroscopy 
of isotopically identified fission fragments with AGATA and VAMOS++" }

\maketitle

\abstract{
$\gamma$-ray spectroscopy of fission fragments is a powerful method for studies of nuclear structure 
properties. Recent results on the spectroscopy of fission fragments, 
using the combination of the AGATA $\gamma$-ray tracking array and the VAMOS++ 
large acceptance magnetic spectrometer at GANIL, are reported.  
A comparison of the  performance of the large germanium detector arrays 
EXOGAM and AGATA illustrates the advances in $\gamma$-ray spectroscopy of fission fragments.
Selected results are highlighted for prompt $\gamma$-ray spectroscopy studies, 
measurements of short lifetimes of excited states with the Recoil Distance Doppler-Shift method,
using both AGATA and VAMOS++ and prompt-delayed $\gamma$-ray spectroscopy studies using AGATA, VAMOS++ and EXOGAM.
}
%

%
\setcounter{secnumdepth}{2} \setcounter{tocdepth}{2}

\section{Introduction}

\label{intro}

Nuclear fission is one of the most effective ways of producing and studying 
neutron-rich exotic isotopes. Fission fragments cover a wide range of the nuclear 
chart and exhibit a variety of 
phenomena ranging from single-particle excitations, near shell closures, to collective 
excitations related to nuclear vibrations 
or deformations. The $\gamma$-ray spectroscopy of fission fragments can be used to probe 
the evolution of nuclear structure properties 
as a function of excitation energy, angular momentum and neutron-proton
asymmetry~\cite{Hamilton1995,Ahmad1995,Navin2014, Nav2014Yb,Leoni2021}.

The prompt $\gamma$-rays emitted by the secondary fission fragments, as they de-excite to their 
ground states, provide detailed insight into 
the structure of nuclei at large spin and isospin. The prompt $\gamma$-rays are emitted on a very 
short time scale {(less than few nanoseconds)}, after scission, although sometimes isomers can delay the decay 
process~\cite{Metag1980}. The study of prompt $\gamma$ rays faces the challenge of identifying a
particular $\gamma$-ray transition among all $\gamma$ rays emitted by few hundred of fission 
fragments produced in a single experiment. 
The use of known characteristic $\gamma$ rays, in the fragment of interest or in the complementary
partner fragment, has been proven to be a 
powerful tool for characterisation of fission fragments~\cite{Hamilton1995,Ahmad1995}. 
Experiments making use of high-fold $\gamma$-ray coincidence techniques, involving the
Gammasphere~\cite{Lee1997}, EUROGAM~2~\cite{Urban1997} and EUROBALL~\citep{EUROBALL} arrays, 
to study fission fragments produced
in either spontaneous-fission process or in in-beam heavy-ion 
induced fission reactions using stable beams, were used to cover a broad range of topics in nuclear
structure~\cite{Hamilton1995}. 
More recently, a similar approach was used in conjunction with fission induced by cold and fast neutrons~\cite{Jentschel_2017,Lebois2020}.

\begin{table*}[!ht]
\caption{Summary of the main characteristics of the fission fragment spectroscopy experiments 
performed with VAMOS++ and AGATA. The beam was $^{238}$U at 6.2MeV/A in all the cases. }
{
\label{tab:ExpSummary}
\begin{center}
\begin{tabular}{|l|ll|llll|lll|l|}
\hline
     & & &  \multicolumn{4}{l|}{VAMOS}     & \multicolumn{3}{l|}{AGATA}   & \multicolumn{1}{l|}{References}                                                                                                                \\ \hline

Exp.~   & \multicolumn{1}{c|}{Duration}  & \multicolumn{1}{c|}{VAMOS-$\gamma$} &
\multicolumn{1}{c|}{Ang.} & \multicolumn{1}{c|}{B$\rho$} & \multicolumn{1}{c|}{Tar./Thick.} & \multicolumn{1}{c|}{Deg./Thick.} &  
\multicolumn{1}{c|}{Dist.} & \multicolumn{1}{c|}{Crystals} & \multicolumn{1}{c|}{Ang. Range}   & 
\multicolumn{1}{l|}{}   \\ 
   & \multicolumn{1}{c|}{[h]} & \multicolumn{1}{c|}{[events]} & 
\multicolumn{1}{c|}{[deg.]} & \multicolumn{1}{c|}{[Tm]}  & \multicolumn{1}{c|}{[$\mu$m]} & \multicolumn{1}{c|}{[$\mu$m]} &  
\multicolumn{1}{c|}{[cm]} & \multicolumn{1}{c|}{} & \multicolumn{1}{c|}{[deg.]}  & 
\multicolumn{1}{l|}{}   \\ 

\hline
1    & \multicolumn{1}{c|}{336} & \multicolumn{1}{c|}{$4\times 10^8$} &\multicolumn{1}{l|}{20}      & \multicolumn{1}{l|}{1.1}     & \multicolumn{1}{l|}{$^9$Be/1.6, 5 }  & \multicolumn{1}{l|}{-}  & 
 \multicolumn{1}{l|}{13.5}            & \multicolumn{1}{c|}{32} & \multicolumn{1}{l|}{90-170}  & 
\multicolumn{1}{l|}{\cite{Kim2017,Biswas2019,Biswas2020,Banik2020a}}   \\ \hline
2    &  \multicolumn{1}{c|}{336} & \multicolumn{1}{c|}{$6\times 10^8$} & \multicolumn{1}{l|}{28}      & \multicolumn{1}{l|}{1.1}     & \multicolumn{1}{l|}{$^9$Be/10}  & \multicolumn{1}{l|}{-}  &  
\multicolumn{1}{l|}{13.3}           & \multicolumn{1}{c|}{24} & \multicolumn{1}{l|}{90-170} & 
\multicolumn{1}{l|}{\cite{Dudouet2017, Dudouet2019,Rezynkina2022}}   \\ \hline
3    &  \multicolumn{1}{c|}{248} & \multicolumn{1}{c|}{$2\times 10^8$} & \multicolumn{1}{l|}{28}      & \multicolumn{1}{l|}{1.1}     & \multicolumn{1}{l|}{$^9$Be/11}  & \multicolumn{1}{l|}{ Mg/18.5}  &  
\multicolumn{1}{l|}{18.6}           & \multicolumn{1}{c|}{24} & \multicolumn{1}{l|}{120-170}  &
\multicolumn{1}{l|}{\cite{Delafosse2018, Delafosse2019}}   \\ \hline
4    &  \multicolumn{1}{c|}{248} & \multicolumn{1}{c|}{$4\times 10^8$}  &\multicolumn{1}{l|}{19}      & \multicolumn{1}{l|}{1.11}     & \multicolumn{1}{l|}{$^9$Be/10 }   & \multicolumn{1}{l|}{Mg/16.7}    &    
 \multicolumn{1}{l|}{23.5}          & \multicolumn{1}{c|}{35}  & \multicolumn{1}{l|}{120-170} &
 \multicolumn{1}{l|}{\cite{PhDAnsari}}   \\ \hline
\end{tabular}

\end{center}
}
\end{table*}

The necessity of knowledge of the characteristic $\gamma$ rays could be overcome by employing the isotopic identification techniques of fission
fragments using large acceptance magnetic spectrometers such as VAMOS++~\cite{Rejmund2011} and PRISMA~\cite{Montagnoli2005}. 
{ The use of fission induced by reaction in inverse kinematics in conjunction with these large acceptance spectrometers resulted in an higher detection efficiency. This combination has opened new opportunities  to study prompt and delayed $\gamma$ rays emitted by fission fragments~\cite{Navin2014,Nav2014Yb,Rejmund2015a,Rejmund2016}.}
The use  of the VAMOS++ spectrometer with the EXOGAM~\cite{Sim00} large $\gamma$-ray array allowed the 
first assignment of prompt 
$\gamma$ rays in several members of the isotopic chains of Ag~\cite{Kim2017a}, Rh~\cite{Nav2016}, 
Cd and In~\cite{Rejmund2015a,Rejmund2016}. The combination of prompt $\gamma$-ray data-sets 
obtained in coincidence with the VAMOS++ magnetic spectrometers 
with those obtained using high fold $\gamma$-ray techniques using Gammasphere allowed extended 
studies in the isotopic chains of Y~\cite{Wang2021}, Pr~\cite{Wang2015} 
and Pm~\cite{Bhattacharyya2018}, which illustrate the complementarity of both 
methods. Furthermore, the experiments with VAMOS++ 
and EXOGAM using the Recoil Distance Doppler-Shift Method (RDDS)~\cite{Dewald2012} 
allowed the measurement of lifetimes of excited states in isotopes of Zr~\cite{Singh2018}, 
Y and Nb~\cite{Hagen2017} and Tc and Rh~\cite{Hagen2018}.

The advent of the new generation of $\gamma$-ray tracking arrays AGATA~\cite{AGATA} 
and GRETINA~\cite{Paschalis2013} allows an improved
determination of the spatial position of the first interaction point of each $\gamma$-ray 
in the detector and an increase of the operating counting rate 
with larger $\gamma$-ray multiplicities. Thus the increased effective granularity results 
in an improved Doppler correction capability of the energy of the 
$\gamma$-rays emitted by nuclei in flight, provided that the velocity vector 
$\overrightarrow{\rm{v}}$ of the recoiling fragment is measured on an event-by-event basis 
with sufficient precision. To ensure that the final Doppler-corrected $\gamma$-ray energy
resolution only arises from the $\gamma$-ray tracking capabilities,
resolution in the scattering angle of the fragment better than
$1^\circ$ and resolution in the interaction point at the target than
1~mm, are required.

Furthermore, the continuous angular coverage of $\gamma$-ray tracking arrays provides new opportunities for lifetime measurement in the picosecond range based on 
Doppler-Shift method~\cite{Stahl2017}. 

This paper presents recent results on the spectroscopy of fission fragments 
using the AGATA $\gamma$-ray tracking array combined with the VAMOS++
large acceptance spectrometer at GANIL~\cite{Clement2017}.  
The presented data arise from four experiments (whose main characteristics
are described in Table~\ref{tab:ExpSummary}) that can be summarised as follows:
\begin{description}
\item[- \bf{Exp.~1}:] Prompt-delayed $\gamma$-ray spectroscopy of 
$^{122-131}$Sb \cite{Biswas2019},
$^{119-121}$In~\cite{Biswas2020}, $^{130-134}$I~\cite{Banik2020a} and experimental methods~\cite{Kim2017},  
\item[- \bf{Exp.~2}:] Prompt $\gamma$-ray spectroscopy of $^{96}$Kr~\cite{Dudouet2017} and  $^{81}$Ga~\cite{Dudouet2019},
$^{83,85,87}${As}~\cite{Rezynkina2022},
\item[- \bf{Exp.~3}:] Lifetime measurements in $^{84}$Ge, $^{88}$Kr, 
$^{86}$Se~\cite{Delafosse2018, Delafosse2019},
\item[- \bf{Exp.~4}:] Lifetime measurements in neutron-rich Zr, Mo and Ru~\cite{PhDAnsari}.
\end{description}

\begin{figure*}[!ht]
\begin{center}

  \includegraphics[width=\textwidth]{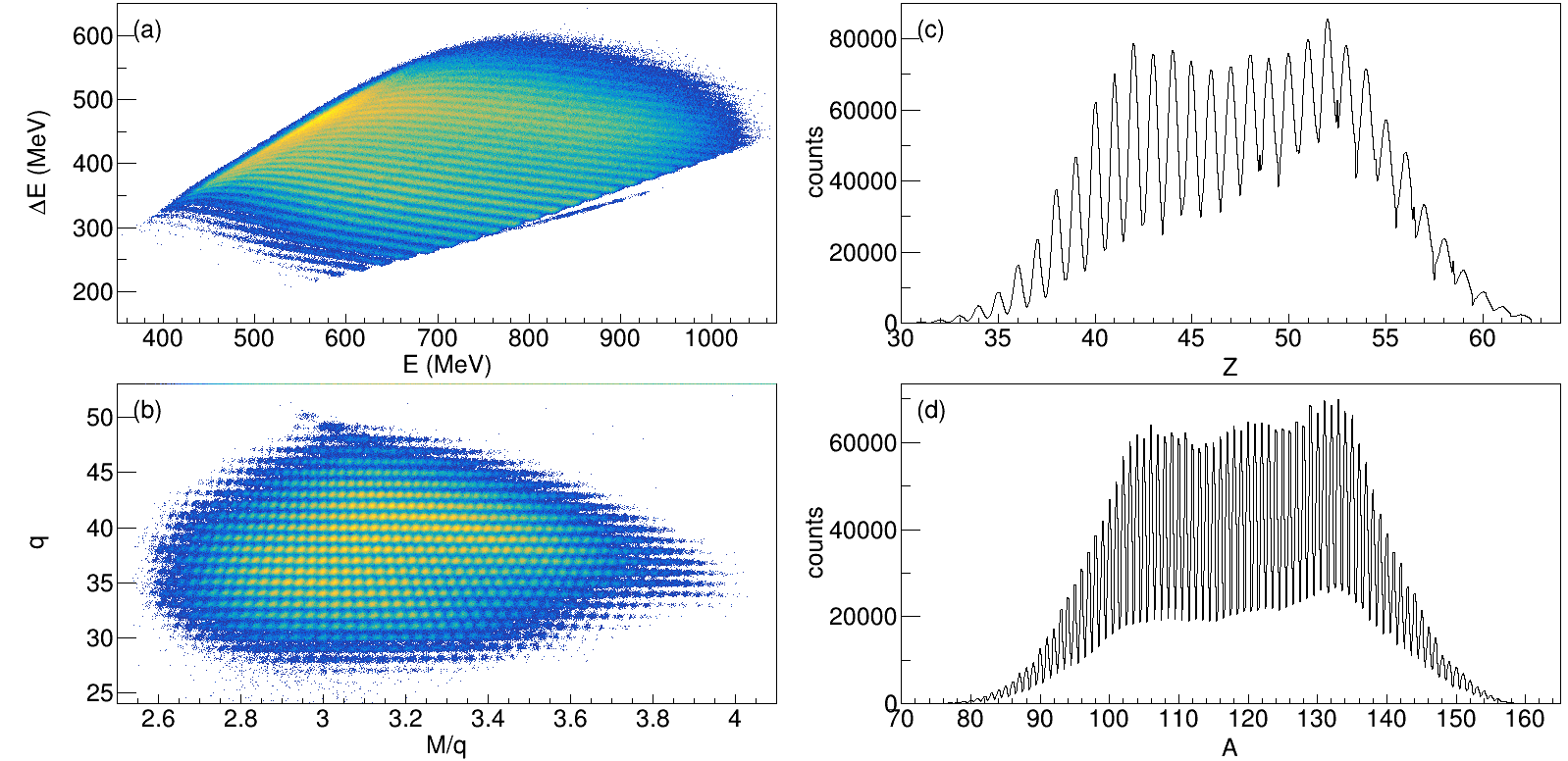} 

\end{center}
\caption{VAMOS identification spectra for fission fragments produced in the $^{238}$U+$^{9}$Be reaction at 6.2 MeV/A (a) two-dimensional spectra of energy loss ($\Delta$E) as function of 
total energy $E$ measured in ionisation chambers of VAMOS++ 
(b) two-dimensional spectra of the atomic number $Z$ as a function of the mass-over-charge ratio ($M/q$) 
(c) atomic number ($Z$) distribution (d) atomic mass distribution (A). The data is taken from the 
Exp.~1, see Table~\ref{tab:ExpSummary}.}
\label{fig:1}
\end{figure*}

The performance of VAMOS++ for the isotopic identification of fission fragments produced in inverse kinematics is presented in Sec.~\ref{sec:SecID}. 
The Doppler correction of the $\gamma$-ray energy is discussed in Sec.~\ref{sec:SecDop}.
A comparison of performances for the spectroscopy of fission fragments
between EXOGAM and AGATA is used to illustrate advances in fission fragment spectroscopy. 
Recent results for prompt $\gamma$-ray spectroscopy are highlighted in Sec.~\ref{sec:PrGamSp} and for the measurement of lifetimes of excited states in Sec.~\ref{sec:RDDS}. Finally, Sec.~\ref{sec:PrDelSp} presents results from prompt-delayed $\gamma$-ray spectroscopy with combinations of AGATA and EXOGAM with VAMOS++. 

\section{Isotopic Identification of Fission Fragments}
\label{sec:SecID}
The fission fragments were typically produced in fusion and transfer induced fission 
by a  $^{238}$U beam at the energy of $6.2$~MeV/A on 
a $^9$Be target of typical thickness ranging from of $1.6~\rm{\mu  m}$ to $10~\rm{\mu  m}$.  
A typical beam intensity of $\sim 1~\rm{pnA}$ was used.  
Fission fragments were isotopically identified in terms of atomic number $Z$, mass number $A$ 
and atomic charge $q$, in the VAMOS++ spectrometer, placed at angles between $20$ and $28$
degrees depending on the fragments of interest. { One of the two emitted fragments is
  detected and isotopically identified in the VAMOS++.} 
The VAMOS++ focal plane detection system consisted of a multi-wire proportional counter (MWPC) 
(stop of the time-of-flight of the ion), two drift chambers (horizontal and vertical 
tracking of the fragment trajectory, $X, \theta, Y, \phi$) and an segmented ionisation chamber 
(energy loss and energy of the ion, $\Delta  E, E)$.   
The ionisation chamber was filled with CF${_4}$ gas at pressures between $70-100~\rm{mbar}$, depending on the ions of interest.  
A dual position sensitive MWPC~\cite{Vandebrouck2016} (DPS-MWPC) (start of the time-of-flight, 
horizontal and vertical tracking of the fragment trajectory, $\theta_{target}, \phi_{target}$) was placed at the entrance of the spectrometer. The MWPCs and drift chambers were filled 
with isobutane gas at a pressure of $6~\rm{mbar}$. The fission fragments were 
implanted in the gas inside the ionisation chamber. The atomic number $Z$ of the ions was obtained based on $\Delta E-E$ correlation technique. 
The mass number $A$ was obtained from the reconstructed magnetic rigidity, flight path and the measured time-of-flight. { Details 
on the identification techniques and performances can be found in  Ref.~\cite{Rejmund2011} while details on the acceptance of the spectrometer for fission reactions are described in Ref.~\cite{Navin2014}}. { Typical fission fragment rates in the VAMOS++ focal plane were ranging between $5$ and $10$kHz and were limited by the pileup in the drift chambers and ionization chambers.}

In Fig.~\ref{fig:1} the typical identification spectra obtained for fission fragments using VAMOS++
are shown. The two-dimensional correlations are shown in panel (a) energy loss versus total energy ($\Delta E$ vs. $E$) and (b) atomic charge versus mass-over-charge ($q$ vs. $M/q$).
The corresponding one-dimensional spectra are shown in panel (c) atomic number ($Z$) and (d)
atomic mass ($A$). The data is taken from the 
Exp.~1, see Table~\ref{tab:ExpSummary}.

\begin{figure}[t]
\centering

  \includegraphics[width=\columnwidth]{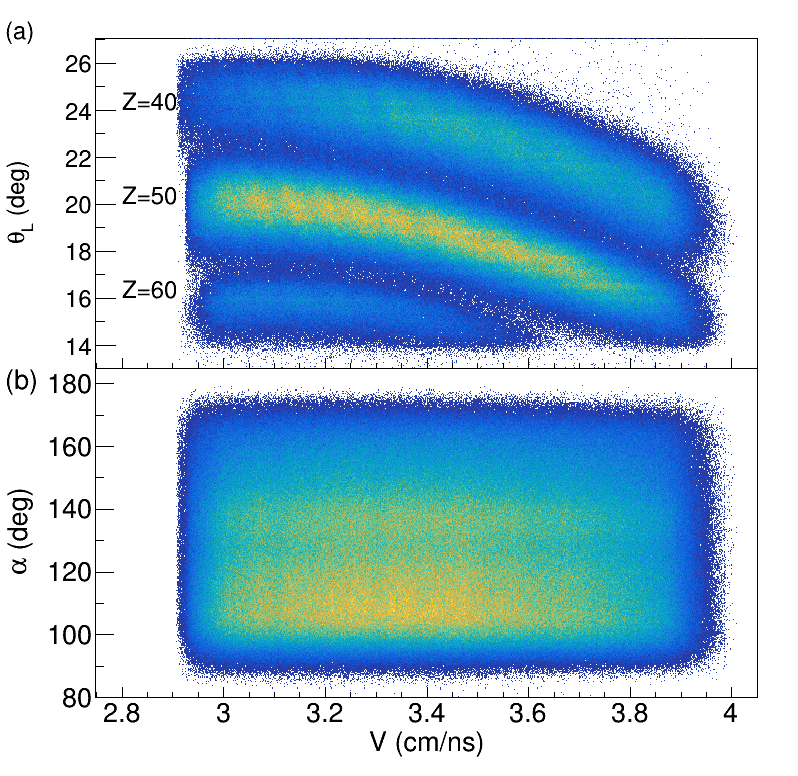}
\caption{(a) The laboratory angle $\theta_{L}$ versus the velocity
  $\rm{v}$ of fission fragments with the atomic number $Z=40$, $50$
  and $60$.  (b) The angle $\alpha$, between the $\gamma$-ray emission
  vector $\overrightarrow{\rm{v_\gamma}}$ and the velocity vector
  $\overrightarrow{\rm{v}}$ (see also Sec.~\ref{sec:SecDop}), versus
  the velocity $\rm{v}$ of fission fragments for the same dataset as
  pannel (a).  The data is taken from the Exp.~1, see
  Table~\ref{tab:ExpSummary}. }
\label{fig:2}
\end{figure}

The velocity vector $\overrightarrow{\rm{v}}$  
of the fragment was  measured  using  the  DPS-MWPC  detector as  described  in Ref.~\cite{Vandebrouck2016}. Figure~\ref{fig:2}(a) 
shows the correlation between 
the angle of the fragment in the laboratory system ($\theta_L$) detected in VAMOS++
and its velocity $\rm{v}$ for fission fragments with the atomic number $Z=40$, $50$ and $60$. The data is taken from the Exp.~1, see Table~\ref{tab:ExpSummary}.
The typical portion of the kinematics of fission fragments measured can be seen in the figure. 
The much stronger kinematic focusing of the heavier fragments with respect to the lighter ones 
can be seen. Figure~\ref{fig:2}(b) shows
the correlation of the angle $\alpha$, between the $\gamma$-ray emission vector 
$\overrightarrow{\rm{v_\gamma}}$ versus the velocity vector $\overrightarrow{\rm{v}}$ { of the detected fragment} 
(see also Sec.~\ref{sec:SecDop}), and the velocity $\rm{v}$.
In Fig.~\ref{fig:2} one can observe the angular opening of VAMOS++ (panel (a)) and AGATA (panel (b)).
Also, from the range of the velocity $\rm{v}$ of $2.9 - 4$~cm/ns and the mean flight path in VAMOS
$D=760$~cm, one can infer the typical time-of-flight from the target to the focal plane of 
$190 - 260$~ns.

\section{Doppler Correction of $\gamma$ ray energy}
\label{sec:SecDop}
The prompt $\gamma$ rays ($\gamma_P$), emitted near the target position were detected 
by the AGATA~\cite{AGATA} $\gamma$-ray tracking array { and acquired in coincidence with fragment detected in VAMOS.}
The array was placed at a distances from the target ranging from $13.5$~cm to $23.5$~cm
depending on the configuration used, see Ref.~\cite{AGATA, Clement2017} for details of the different configurations.
{ The detection efficiency of the AGATA array in the different configurations is discussed in Refs.~\cite{Clement2017,LjungVall2020, Kim2017}.}
{ Typical counting rates in the AGATA detectors was $\sim 20$~kHz per crystal. }
The AGATA array holding structure and the VAMOS++ spectrometer were supported on a common platform 
which could rotate around a vertical axis perpendicular to the beam axis at the
target position.
The AGATA detectors typically covered angles from $\sim 100^\circ$ to $\sim 170^\circ$, relative to the axis of the VAMOS++ spectrometer. 
The $\gamma$-ray emission vector $\overrightarrow{\rm{v_\gamma}}$ was determined using the first
three-dimensional interaction point of the $\gamma$-ray in the AGATA array, obtained from pulse shape analysis and tracking 
procedures~\cite{AGATA}. A typical position resolution of 
$\sim 5$~mm~(FWHM)~\cite{PhDRecchia,  Recchia2009, Soderstrom2011, LjungVall2020} has been 
reported for $\gamma$-ray energies around $1.3$~MeV. The measured velocity vector of the { detected fragment}
$\overrightarrow{\rm{v}}$ and $\gamma$-ray emission vector $\overrightarrow{\rm{v_\gamma}}$ were 
used to derive the Doppler-corrected $\gamma$-ray energy on an event-by-event basis.
The Doppler correction was obtained using the following relationship
$$E_{RF} = E_{LAB} \cdot (1-\beta \cdot cos(\alpha)) \cdot \gamma$$
where: $E_{RF}$ and $E_{LAB}$ are respectively the energies of the
$\gamma$ ray in the rest frame of the nucleus and in the laboratory
system, $\beta = v/c$, $\gamma = \frac{1}{\sqrt{1 - \beta^2}}$ and
$\alpha$ is the angle between vector $\overrightarrow{\rm{v}}$ and
$\overrightarrow{\rm{v_\gamma}}$. { It should be noted that
  $\gamma$ rays emitted by the complementary fragment will have an
  incorrect doppler correction applied, resulting in Doppler broadened
  and shifted peaks that contribute to the background of the
  spectra. It was demonstrated in Refs.~\cite{Navin2014,Nav2014Yb}
  that Doppler correction for the binary partner can be achieved using
  two-body kinematics derived from the velocity vector of the measured
  fragment.}

\begin{figure}[t]
\centering
 \includegraphics[width=\columnwidth]{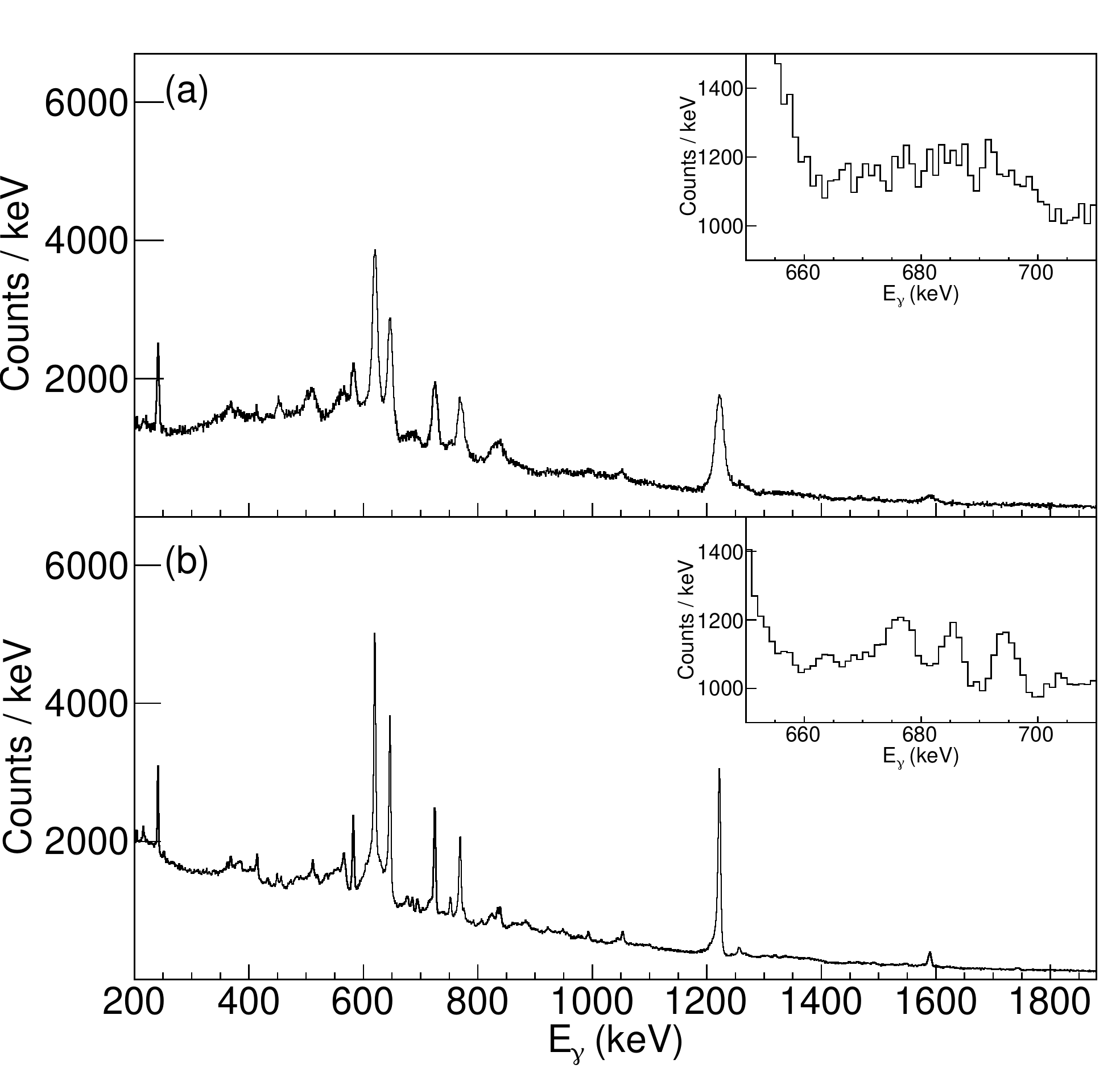}
 \caption{  Prompt Doppler-corrected $\gamma$-ray spectra measured in coincidence with isotopically identified $^{98}$Zr in VAMOS++:  (a) with $\gamma$ rays measured with the EXOGAM array (b) with $\gamma$ rays measured with AGATA $\gamma$-ray tracking array. The two $\gamma$-ray spectra are normalised to the same total number of counts. The insets highlight part of the corresponding $\gamma$-ray spectra in the $650$~keV to $710$~keV range.}
\label{fig:3}       
\end{figure}
\section{Prompt-$\gamma$-ray spectroscopy
}
\label{sec:PrGamSp}

\subsection{AGATA versus EXOGAM}

To illustrate the performances obtained with AGATA for the prompt-$\gamma$-ray 
spectroscopy of fission fragments, the Fig.~\ref{fig:3} shows 
Doppler-corrected $\gamma$-ray spectra in coincidence with isotopically identified 
$^{98}$Zr in VAMOS++. Figure~\ref{fig:3}(a) shows the prompt  
$\gamma$ rays measured with the EXOGAM array~\cite{Sim00} consisting of 
$11$~Compton-suppressed segmented clover HPGe detectors ($15$~cm away from the target),
in coincidence with the isotopically identified fragments. 
The $\overrightarrow{\rm{v}}$ of the fragment  
along with the position of the center of the electrical segment of the clover detector that had registered the
highest energy deposit were used to obtain the $\gamma$-ray energy   
in the rest frame of the emitting fragment~\cite{Sam08}. 
Figure~\ref{fig:3}(b) shows the prompt $\gamma$ rays measured with the AGATA array 
in coincidence with the isotopically identified fragments. The data is taken from Exp.~1, see Table~\ref{tab:ExpSummary}.

The comparison of the spectra clearly illustrates the improved $\gamma$-ray energy resolution arising 
from the position resolution of the first interaction point derived in AGATA. 
The $1222.9$~keV $\gamma$ ray, $2^+ \rightarrow 0^+$  transition in $^{98}$Zr,
can be used to evaluate the obtained energy resolution of Doppler-corrected spectra. 
Considering all clovers from EXOGAM (including $90^\circ$ and 
$135^\circ$ rings), a resolution of $15$~keV was obtained. Considering only backward angles in EXOGAM, a resolution of $7$~keV was measured. This is to be 
compared with a resolution of $5$~keV obtained with AGATA for angular coverage ranging from $100^\circ$ to $170^\circ$. The improved resolving power can be further seen in the insets of  Fig.~\ref{fig:3}, where within an expanded region of the $\gamma$-ray 
spectra several weak transitions, unresolved using the EXOGAM array, could be resolved 
using the AGATA array.

\subsection{$\gamma$-ray spectroscopy of $^{96}$Kr}
\label{sec:96Kr}
\begin{figure}[t]

\centering
  \includegraphics[width=0.9\columnwidth]{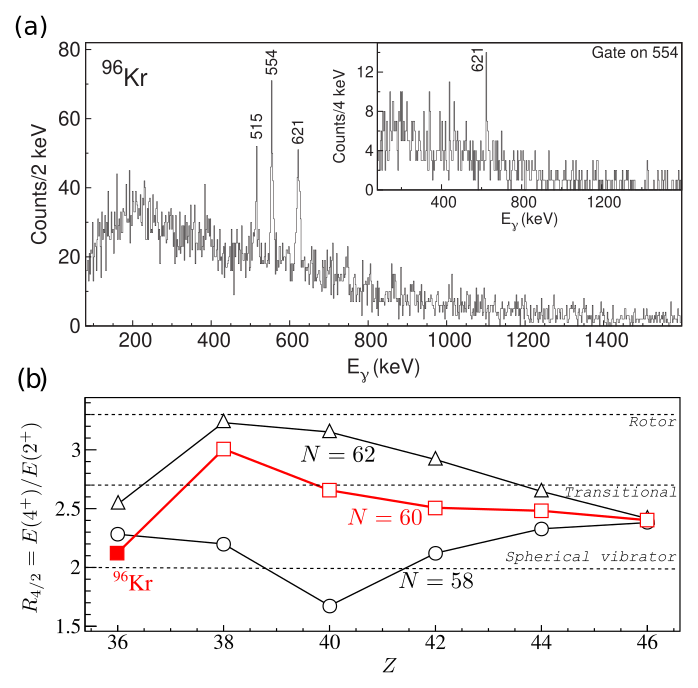}

\caption{
The data is obtained from the Exp.~2, see Table~\ref{tab:ExpSummary}. 
(a) Doppler-corrected $\gamma$-ray spectra measured in AGATA in coincidence with $^{96}$Kr detected and identified isotopically in VAMOS++~\cite{Dudouet2017}.
(Inset) $\gamma$-ray spectrum measured in coincidence with the $554$~keV transition.
(b) The $R_{4/2}$ ratio against the atomic number $Z$ for $N = 58–62$ isotonic chains. 
Horizontal dashed lines represent a schematic classification between spherical
vibrational and rotor nuclei from Ref.~\cite{Casten1990}. The $R_{4/2}$ point 
for $^{98}$Kr at $N=62$ obtained from Ref.~\cite{Flavigny2017} has been added to the figure. The figure is adapted from  Ref.~\cite{Dudouet2017}~\href{https://creativecommons.org/licenses/by-sa/4.0/}{CC BY 4.0}. \copyright~2017, J. Dudouet {\it et al.}, published by American Physical Society.}

\label{fig:4}
\end{figure}

The sudden appearance of the onset of the collectivity at $N=60$ has been 
one of the early successes of the $\gamma$-ray spectroscopy of fission fragments. 
After decades of studies establishing the sudden transition towards the deformation
at $N=60$ in Zr ($Z=40$) and Sr ($Z=38$), the detailed description of this island 
of deformation still challenges theoretical models. In Ref.~\cite{Dudouet2017}, 
the prompt $\gamma$-ray spectroscopy of the neutron-rich $^{96}$Kr ($Z=38$ and $N=60$)
has contributed to delineate the limits of this island of deformation. 
The nucleus of interest, $^{96}$Kr, was produced in transfer- and fusion-induced 
fission processes, using the $^{238}$U beam impinging on a $^9$Be target. 
This experiment is referred to as Exp.~2, 
see Table~\ref{tab:ExpSummary}.
The prompt Doppler-corrected $\gamma$-ray spectrum measured in coincidence with 
$^{96}$Kr isotopically identified in VAMOS++ is shown in Fig~\ref{fig:4}(a). 
Three $\gamma$-ray transitions at the energies of  $554(1)$~keV, $621(2)$~keV, 
and $515(2)$~keV can be seen in the spectrum. The $554$~keV transition confirms the
excitation energy of the first $2^+$ state of 
Ref.~\cite{Albers2012}. 
The $621$~keV transition was observed in coincidence with the $554$~keV 
transition as can be seen in the inset in Fig.\ref{fig:4}(a).
It was interpreted as the transition depopulating the first $4^{+}$ excited state 
at the energy of $1175(3)$~keV. 
Because of the limited statistics, it was not possible to obtain a significant
coincidence analysis for the $515$~keV transition which was not placed in 
the level scheme. The presence of low-lying $2^+_2$
excited states in the Kr isotopic chain suggests
the possible assignment of the $515$~keV $\gamma$ ray to the
$2^+_2 \rightarrow 2^+_1$ transition.

Recently, the spectroscopy of $^{96}$Kr from knock-out and inelastic
reactions was reported~\cite{Gerst2022}.  A $888~(16)$~keV state,
decaying by $887^{+24}_ {-23}$~keV and $334~(16)$~keV $\gamma$-ray
transitions, was tentatively assigned to the $2^+_2$ state based on
coincidence arguments. These $\gamma$-ray transitions were not
observed in Ref.~\cite{Dudouet2017} due to the available
statistics. The $515~(2)$~keV transition was also observed and
reported in coincidence with the $2^+->0^+$ transition, but it was
not placed either in the level scheme. The nature of the state
depopulated by the $515$~keV transition remains an open question to be
addressed.

To understand, quantify, and characterise the evolution
of the nuclear structure along isotopic chains, a systematic study
of the energy ratio $R_{4/2}=E(4^+)/E(2^+)$ is often used~\cite{Casten1990}.
The newly measured energy of the $4^+$ state results in the  
energy ratio of $R_{4/2}=2.12(1)$.
In Fig.\ref{fig:4}(b) the $R_{4/2}$ ratio is shown as a function of
atomic number $Z$ for isotonic chains with $N=58$, $60$ and $62$.
It is seen that at the $N=58$ the nuclei between Kr and Pd exhibit very 
little collectivity and are situated between the transitional and 
spherical vibrator regime. For $N=60$ and $62$, the collectivity increases 
with decreasing $Z$, reaching the maximum in Sr. In $^{96}$Kr one observes 
an abrupt decrease of the collectivity. The $R_{4/2}$ obtained for
$^{98}$Kr~\cite{Flavigny2017} follows the same behaviour.
 This new measurement highlights an abrupt transition of the degree of
 collectivity as a function of the proton number at $Z=36$.
 A possible reason for this abrupt transition could be related to the
 insufficiently large amplitude of the proton excitation in 
 the g$_{9/2}$, d$_{5/2}$, and s$_{1/2}$ orbitals to generate strong
 quadrupole correlations or coexistence of competing different shapes.

This measurement established the Kr isotopic chain as the low-$Z$ 
boundary of the island of deformation for $N=60$ isotones. 
The comparison with available theoretical predictions using different 
beyond mean-field approaches shows that these models fail to reproduce the
abrupt transitions at $N=60$ and $Z=36$ and that the precise 
description of the region remains challenging. See Ref.~\cite{Dudouet2017} 
for further details.

\subsection{$\gamma$-ray spectroscopy of $^{81}$Ga}
\label{sec:81Ga}

\begin{figure}[t]
  \centering
  \includegraphics[width=1.\columnwidth]{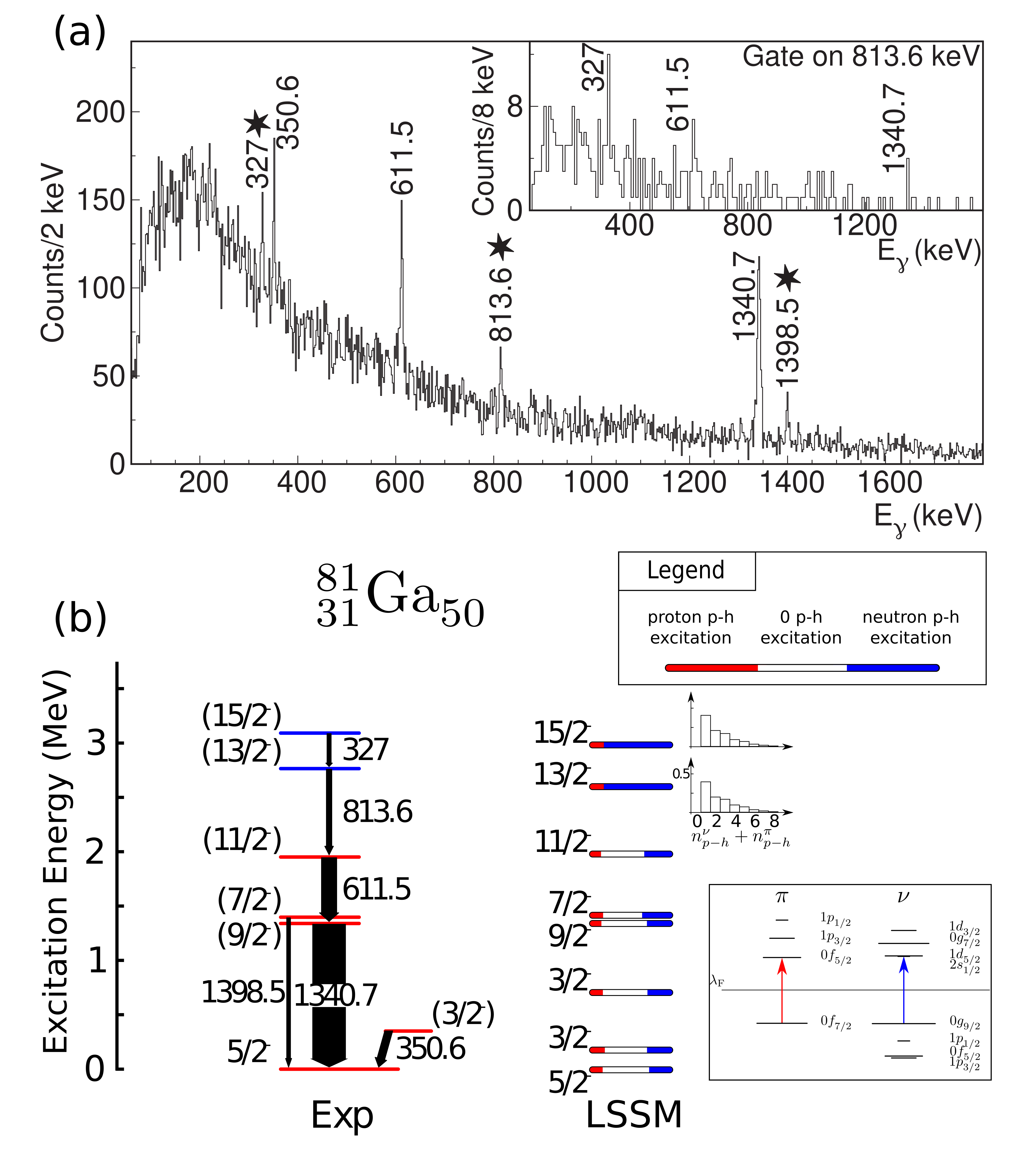}
\caption{(a) Tracked Doppler-corrected $\gamma$-ray spectrum measured in coincidence with 
the isotopically identified $^{81}$Ga. 
Known transitions are labelled by their energy, while newly reported transitions are marked with an additional star symbol. The inset shows the $\gamma-\gamma$
coincidence gated on the $813.6$~keV transition in $^{81}$Ga. (b) Experimental and
theoretical level schemes of $^{81}$Ga. 
The experimental levels marked with red (blue) correspond to 
dominant contribution from $0p-h$ ($1p-h$) excitations of the $^{78}$Ni core. The data is obtained from the Exp.~2, see Table~\ref{tab:ExpSummary}. 
The figure is adapted with permission from Ref.~\cite{Dudouet2019}, \copyright~2019, by American Physical Society.
}
\label{fig:5}
\end{figure}
It is agreed that $^{78}$Ni ($Z=28$, $A=50$) manifests a doubly magic character. 
However, in Ref.~\cite{Nowacki2016} the sudden emergence of collective states and their 
coexistence with the spherical states are predicted. These spherical states
arise mainly from one particle-hole excitations across the magic shell gaps 
($Z = 28$ and $N = 50$). Collective states arise from multi particle-hole 
excitations, giving rise to a deformed collective band, 
providing a striking example of shape coexistence. 
The excited states of $N = 50$ isotones provide complementary insight into the 
coupling of single particle-hole configurations with valence protons where the 
particle-hole configuration are intimately related to the properties of the $N = 50$ 
shell gap.

The high-spin states of the neutron-rich $^{81}$Ga, with three valence protons 
outside a $^{78}$Ni core, were measured for the first time~\cite{Dudouet2019}. 
This experiment is referred to as the Exp.~2, 
see Table~\ref{tab:ExpSummary}.
The tracked Doppler-corrected $\gamma$-ray spectrum obtained in coincidence with 
$^{81}$Ga is shown in Fig.~\ref{fig:5}(a). 
The inset shows the $\gamma$-$\gamma$ coincidence spectrum gated on the 
$813.6$~keV transition. The derived level scheme is shown in  Fig.~\ref{fig:5}(b). 
The newly observed high-spin states in $^{81}$Ga are interpreted using the results 
of state of the art Large Scale Shell Model (LSSM) calculations~\cite{Nowacki2016} 
using the PFSDG-U interaction, see Fig.~\ref{fig:5}(b). 
The lower excitation energy levels are understood as resulting from the recoupling 
of three valence protons to the closed doubly magic core, while the highest 
excitation energy levels correspond to excitations of the magic $N = 50$ neutron core. 

These results support the doubly magic character of $^{78}$Ni and the persistence of the 
$N = 50$ shell closure, but also highlight the presence of strong proton-neutron 
correlations associated with the promotion of neutrons across the magic $N = 50$ 
shell gap, only few nucleons away from $^{78}$Ni. See Ref.~\cite{Dudouet2019} 
for further details.

\section{Lifetime measurement using RDDS method}
\label{sec:RDDS}

\subsection{RDDS method}
\begin{figure}[t]
\centering
  \includegraphics[width=0.9\columnwidth]{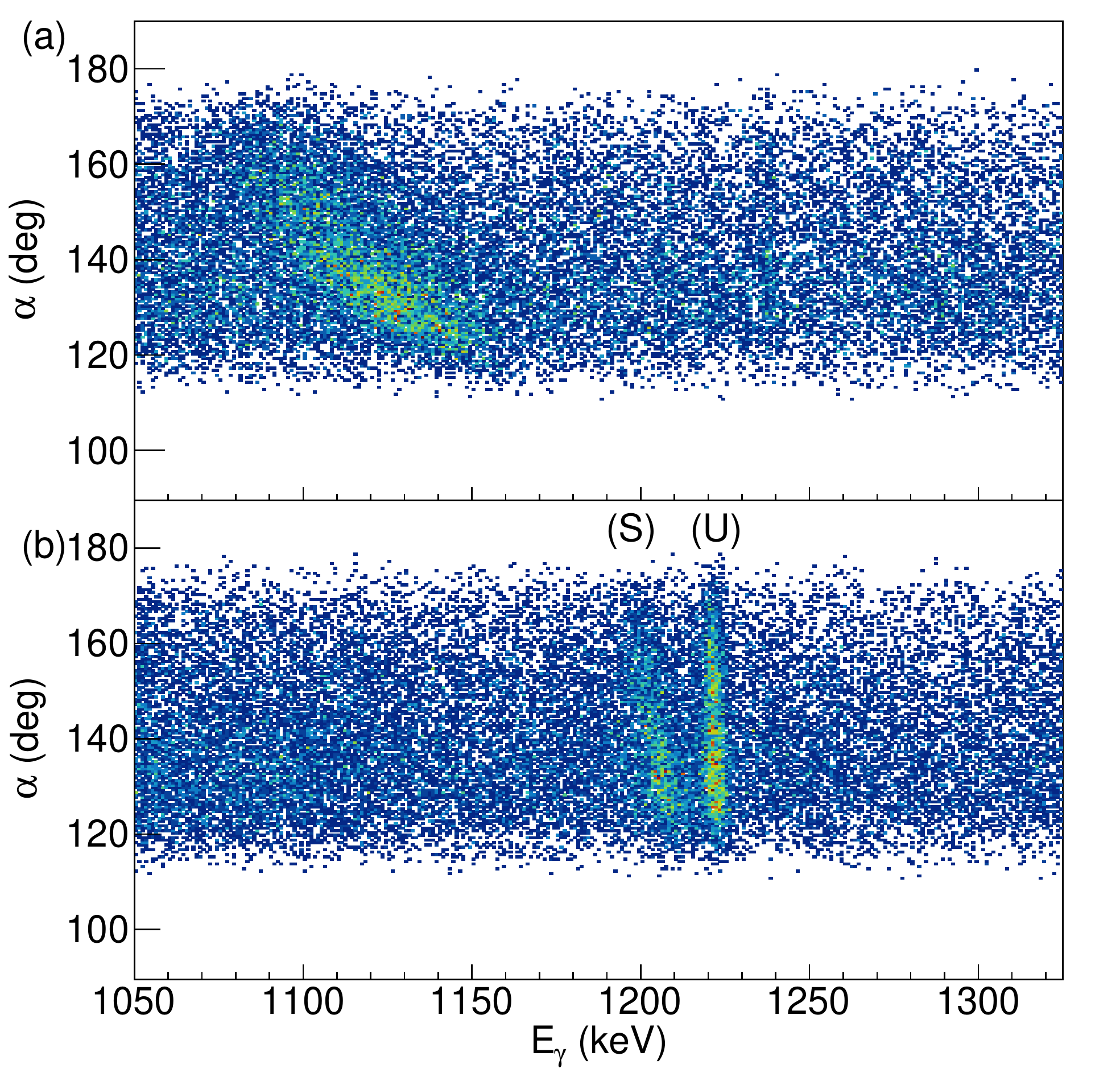}
\caption{Two dimensional spectra of the angle $\alpha$ (the angle between vector $\overrightarrow{\rm{v_d}}$ 
and $\overrightarrow{\rm{v_\gamma}}$) as a function of the $\gamma$-ray energy for a target-to-degrader distance of 470~$\mu$m in coincidence with $^{98}$Zr identified in VAMOS++, (a) detected in the laboratory frame (b) Doppler-corrected using the fragment velocity after the degrader $\overrightarrow{\rm{v_d}}$. The unshifted (U) and Doppler-shifted (S) components are marked. The data is taken from the 
Exp.~4, see Table~\ref{tab:ExpSummary}.}
\label{fig:6}
\end{figure}

\begin{figure*}[ht]
\centering
\includegraphics[width=0.9\textwidth]{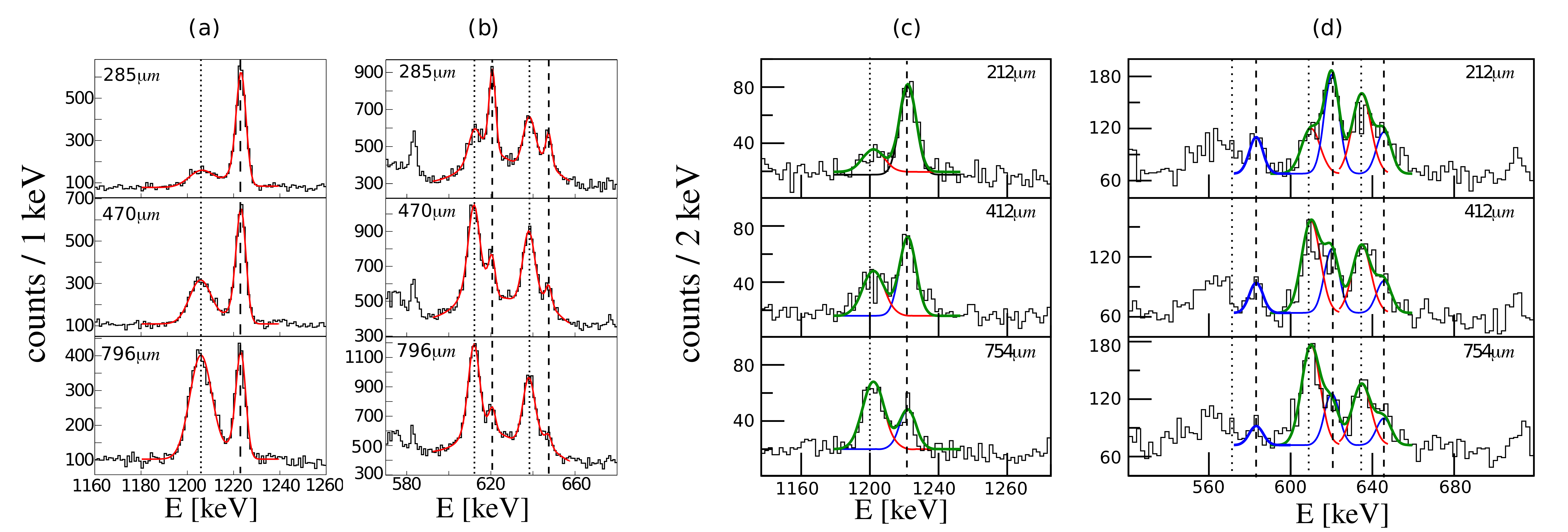}

\caption{Comparison of Doppler-corrected $\gamma$-ray spectra using $\overrightarrow{\rm{v_d}}$, measured in coincidence with isotopically identified $^{98}$Zr in VAMOS++ obtained with
AGATA (a) and (b) and EXOGAM (c) and (d). The target-to-degrader distances are given.
Panels (a) and (c) show Doppler-shifted and unshifted components for the $1222.9$~keV 
transition de-exciting the $2^+$ state. 
Panels (b) and (d) show Doppler-shifted and unshifted components of the $620.5$~keV and 
$647.6$~keV transitions de-exciting the $4^+$ and $6^+$ states.
The data from the AGATA array is taken from the 
Exp.~4, see Table~\ref{tab:ExpSummary} and adapted from Ref.~\cite{PhDAnsari}. The pannel (c) and (d) for the EXOGAM array are adapted with permission from Ref.~\cite{Singh2018},  \copyright~2018, by American Physical Society. 
}
\label{fig:7}
\end{figure*}

The Recoil Distance Doppler-Shift  method is a well established technique for the
determination of picosecond lifetimes of excited nuclear states.
Traditionally, a nucleus produced in a nuclear reaction in a thin target
leaves the target with a velocity $\rm{v_t}$ and is stopped after flying trough a well-defined
distance in a stopper foil. The excited state in the nucleus can de-excite by an emission of 
a $\gamma$ ray either in-flight or at rest in the stopper foil. One can observe,
the intensity of either Doppler-shifted (S) or unshifted (U) components of 
the $\gamma$-ray transition, respectively. The lifetime of the corresponding state
can be measured using the so called decay curve or flight curve based on the
intensities of the Doppler-shifted and unshifted components. 
An alternative procedure called the Differential Decay Curve Method (DDCM) is also used.
See further details in Ref.~\cite{Dewald2012}.

For the experiments, where the recoiling nuclei of interest are to be detected,
as it is the case when using VAMOS++ spectrometer,
the stopper foil can be replaced by a degrader foil. One can observe 
either the $\gamma$ rays emitted after the target at the recoil velocity 
$\rm{v_t}$ or after the degrader foil at $\rm{v_d}$. Typically, the Doppler correction
uses the velocity of the nucleus measured after the degrader $\rm{v_d}$, therefore 
the $\gamma$ rays emitted after the degrader are seen as unshifted and those emitted 
 before the degrader as Doppler-shifted.

In Fig.~\ref{fig:6}(a), the $\gamma$-ray energy measured in the laboratory frame is shown as function of the angle $\alpha$, between the vectors $\overrightarrow{\rm{v_d}}$ and $\overrightarrow{\rm{v_\gamma}}$,
in coincidence with isotopically identified $^{98}$Zr in VAMOS++ for the target-to-degrader distance of $470$~$\mu$m in the Exp.~4, 
see Table~\ref{tab:ExpSummary}.
In Fig.~\ref{fig:6}(b) the same events are shown, but a Doppler correction on an 
event-by-event basis was applied using the measured velocity vector after the 
degrader $\overrightarrow{\rm{v_d}}$. The well-defined in energy, unshifted (U) component
can be seen in the figure.  The Doppler-shifted (S) component becomes dependent on $\alpha$ and appears at a lower energy because the $\gamma$ ray was emitted at a larger velocity, $\rm{v_t} > \rm{v_d}$, and the AGATA array was placed at backward angles, $\alpha>90^\circ$. 

The use of the AGATA array for lifetime measurements using the RDDS method has several 
important assets namely (i) a very good energy resolution for Doppler-corrected 
$\gamma$-ray transitions, (ii) a good coverage for the very backward solid angle where the Doppler effect is largest (iii) the availability of a continuous measurement of the angle $\alpha$, see Fig.~\ref{fig:6}.

\subsection{AGATA versus EXOGAM} 

Doppler-corrected $\gamma$-ray spectra of several transitions in  $^{98}$Zr, isotopically identified in VAMOS++, are shown in Fig.~\ref{fig:7}. Panels (a) and (b) show spectra obtained with AGATA~\cite{PhDAnsari} from Exp.~4, see Table~\ref{tab:ExpSummary}. Panels (c) and (d) show corresponding spectra with comparable target-to-degrader distances obtained with EXOGAM~\cite{Singh2018}. 
Panels (a) and (c) show the components of the $1223$~keV 
transition de-exciting the $2^+$ state ($\tau = 3.8 \pm 0.8$~ps) and panels (b) and (d) the $620$~keV and 
$647$~keV transitions de-exciting the $4^+$ ($\tau = 7.5 \pm 1.5$~ps)  and $6^+$ ($\tau = 2.6 \pm 0.9$~ps)  states. 
The evolution of the intensity of the Doppler-shifted (S) component relative to the unshifted (U) component, for the transitions
de-exciting the states with different lifetimes, is evident as a function of the 
target-to-degrader distance. The improved energy separation between the two components, 
due to the improved first interaction position of the $\gamma$ ray and thus its energy 
resolution can be also seen in the figure. This results in an improved the precision 
of the lifetime analysis and extracted reduced transition strength.

\subsection{Lifetime measurement of excited states in $^{84}$Ge}

\begin{figure}[t]
\centering
 \includegraphics[width=0.9\columnwidth]{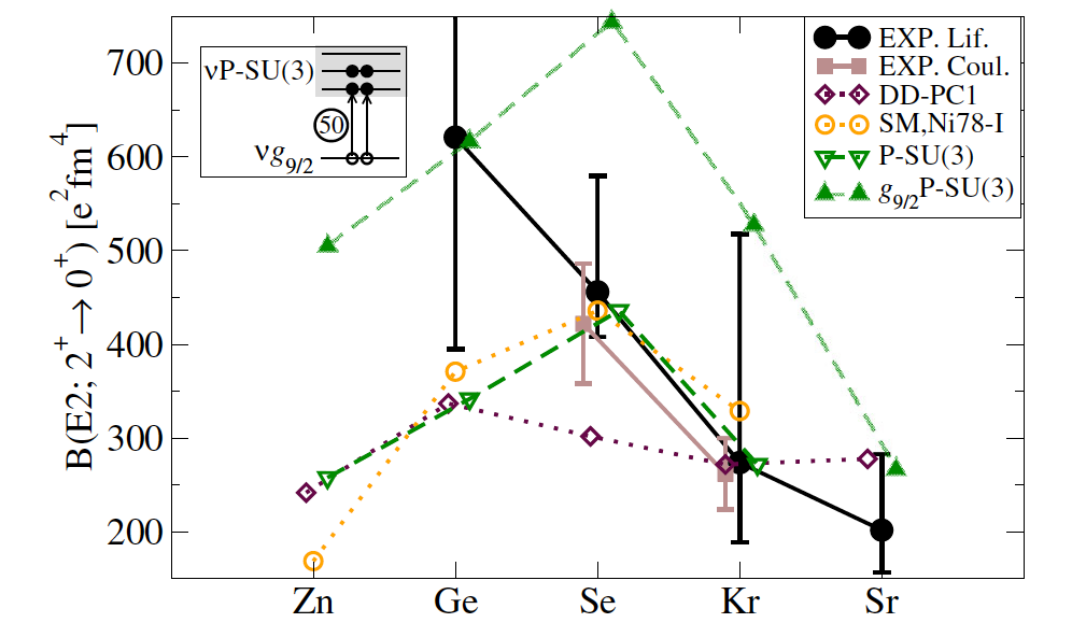}
\caption{
  $B(E2,2^+\rightarrow 0^+)$ systematic of the light $N=52$ even-even isotones from $Z=38$ 
down to $Z=30$. "Exp.~Lif.": experimental values from lifetimes measurements (from Ref.~\cite{Delafosse2018},  except Sr~\cite{Basu2020}; 
 "Exp.~Coul.": experimental values from Coulomb excitation 
 measurements~\cite{Elman2017}; 
 “DD-PC1”: beyond mean- field calculations using the relativistic functional DD-PC1 
 (Ref.~\cite{Delafosse2018}); 
 “SM,Ni78-I” shell-model calculations 
 from~\cite{Sieja2013}; 
“(g$_{9/2}$) P-SU(3)”: pseudo-SU(3) limit~\cite{Sieja2013}  (or including one $N=50$
core-breaking g$_{9/2}$ pair 
promotion, as illustrated by the inset).
The data is obtained from the Exp.~3, see Table~\ref{tab:ExpSummary}. 
(The figures are adapted from  Ref.~\cite{Delafosse2018}~\href{https://creativecommons.org/licenses/by-sa/4.0/}{CC BY 4.0}. \copyright~2018, C. Delafosse {\it et al.}, published by American Physical Society).
}
\label{fig:8}       
\end{figure}

The recent intense experimental and theoretical efforts on the investigation of the nuclear
structure in the vicinity of doubly magic $^{78}$Ni ($Z=28$, $N=50$), have triggered 
experimental measurements of lifetimes of excited states. 

In Ref.~\cite{Delafosse2018}, lifetime measurements of excited states of the light $N=52$ 
isotones $^{88}$Kr ($Z=36)$, 
$^{86}$Se ($Z=34$), and $^{84}$Ge ($Z=32$) using the RDDS method with VAMOS++ and AGATA were 
reported. The reduced electric quadrupole transition probabilities $B(E2,2^+\rightarrow\,0^+)$ 
and $B(E2,4^+\rightarrow\,2^+)$ were obtained for the first time for the hard-to-reach $^{84}$Ge.
The nuclei of interest were produced in transfer- and fusion-induced 
fission processes, using the $^{238}$U beam impinging on a $^9$Be target followed by a Mg 
degrader. This experiment is referred to as the Exp.~3, 
see Table~\ref{tab:ExpSummary}.

Because of low statistics, the RDDS-analysis variant developed in 
Ref.~\cite{Litzinger2015} had 
to be applied, which consists in summing the statistics obtained over all 
distances and determining the lifetime, see Ref.~\cite{Delafosse2018} for further details.

The obtained $B(E2)$ values are placed in the systematics
of light $N=52$ isotones in Fig.~\ref{fig:8}, 
where a comparison with several calculations is also provided. 

Shell-model results from Ref.~\cite{Sieja2013} (open circles), assuming an inert 
$^{78}$Ni core, are in excellent agreement with the experimental values (closed circles) 
obtained for $^{88}$Kr and $^{86}$Se. Interestingly enough, both shell-model and 
experimental values exhaust the limit for pure pseudo-SU(3) symmetry (down triangles) 
for these two isotones. This clearly means that the quadrupole coherence offered by this 
subspace is maximally expressed in these two nuclei. In contrast, both shell-model and 
pseudo-SU(3) values barely reach the lower tip of the experimental error bar for $^{84}$Ge.
The more in-depth analysis of the experimental data in Ref.~\cite{Delafosse2018} suggests,
for the first time, a shape transition from $Z=34$ (soft triaxial) to $Z=32$ (prolate deformed),
a result all the more unexpected as the shell model predicts a "fifth island of inversion" 
only for much lighter ($Z < 28$) systems~\cite{Nowacki2016}.

\section{Prompt-delayed $\gamma$-ray spectroscopy}
\label{sec:PrDelSp}

\subsection{Experimental method}
\begin{figure*}[ht]
\centering
  \includegraphics[width=0.8\textwidth]{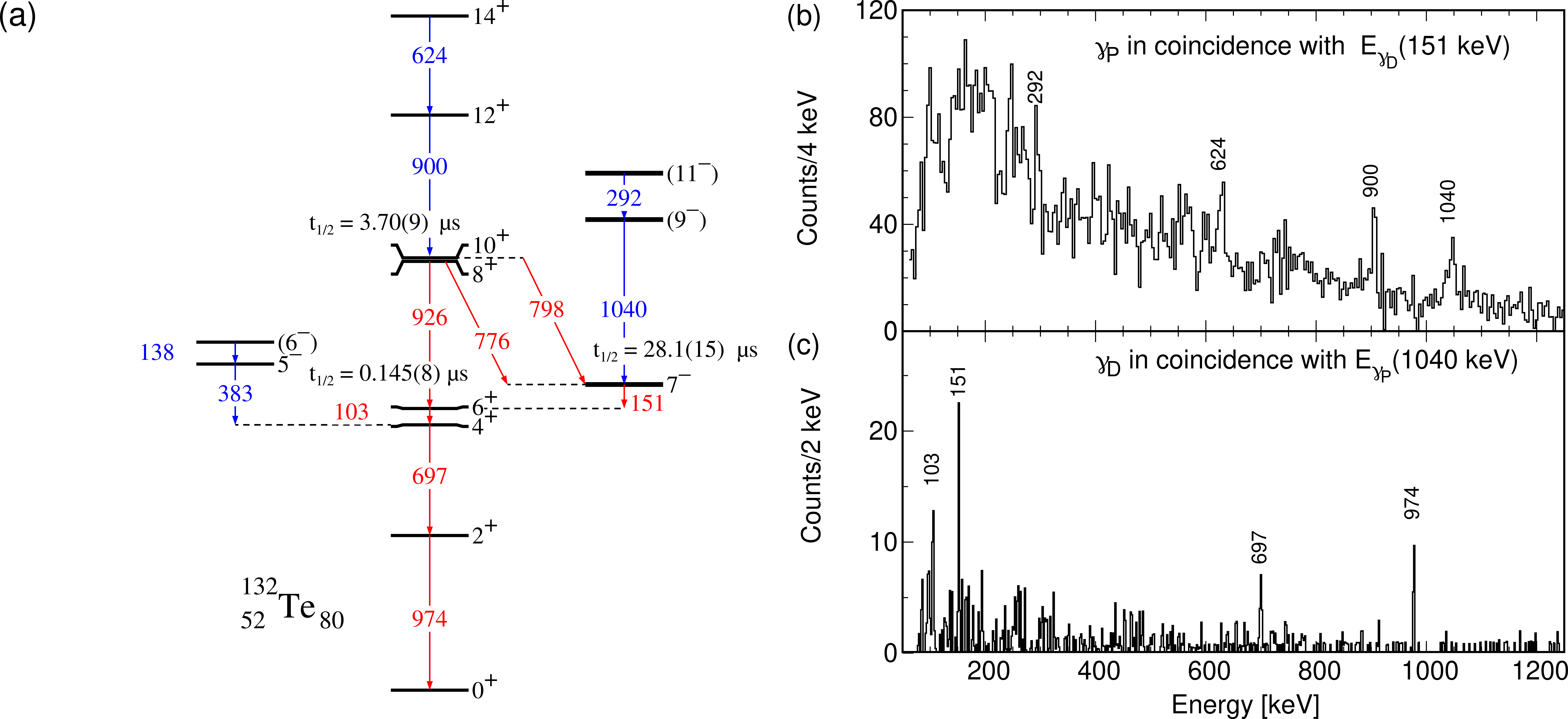}
\caption{
(a) The partial level scheme of $^{132}$Te (below
$4.3$~MeV). The transitions above (below) isomeric states are
indicated in blue (red). 
(b) $^{132}$Te Doppler-corrected prompt $\gamma$ ray ($\gamma_P$) spectrum 
with the condition that a delayed $\gamma$ ray with energy
$E_{\gamma_D} = 151$~keV was detected. 
(c) $^{132}$Te delayed $\gamma$ ray ($\gamma_D$)
spectrum in coincidence with the prompt $E_{\gamma_P} = 1040$~keV $\gamma$ ray.
The data is taken from the Exp.~1, 
see Table~\ref{tab:ExpSummary}. The figures reproduced with permission from Ref.~\cite{Kim2017}, \copyright~2017, published by Springer. 
}
\label{fig:9}
\end{figure*}

\begin{figure*}[ht]
\centering
  \includegraphics[width=0.9\textwidth]{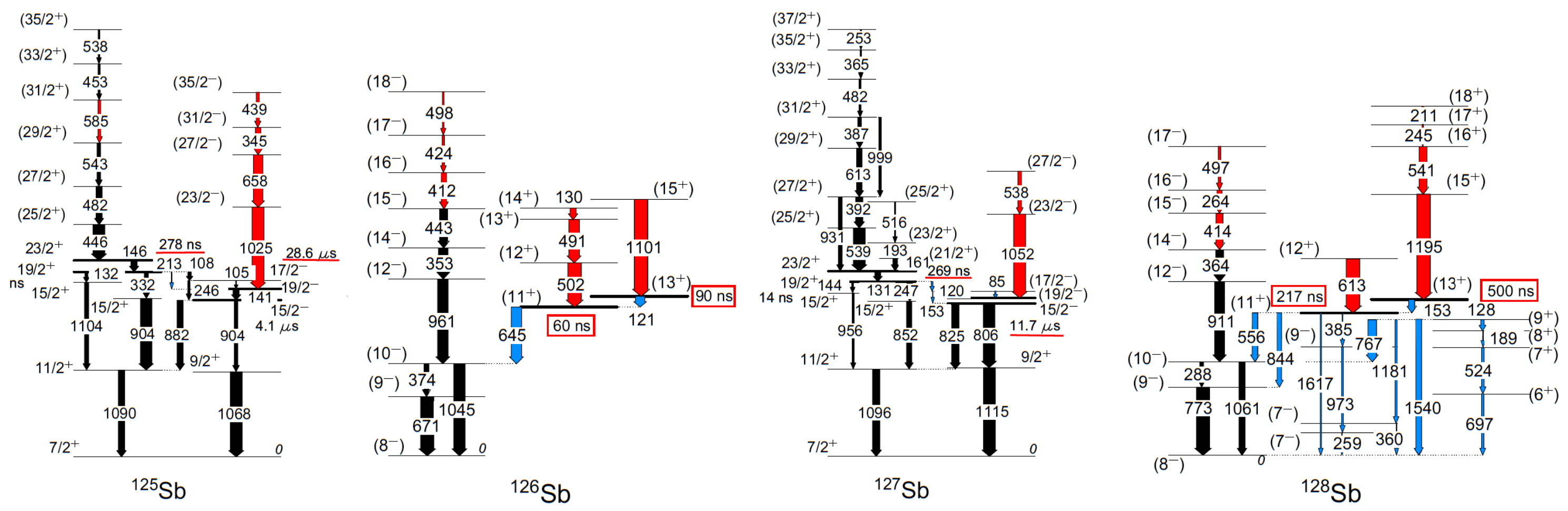}

\caption{
The level schemes of $^{125–128}$Sb. The newly observed $\gamma$-ray transitions above and below the isomer are indicated in red and blue, respectively. The width of the arrows represent
the intensity of the transitions. The isomeric states are indicated by a thick line. Previously known half-lives that were remeasured in Ref.~\cite{Biswas2019} have been underlined by a red line, whereas the newly measured half-lives have been marked with a red box. 
The half-lives not measured in Ref.~\cite{Biswas2019} are also shown. 
The data is taken from the Exp.~1, 
see Table~\ref{tab:ExpSummary}. The figure is adapted from  Ref.~\cite{Biswas2019}~\href{https://creativecommons.org/licenses/by-sa/4.0/}{CC BY 4.0}. \copyright~2017, S. Biswas {\it et al.}, published by American Physical Society.
}
\label{fig:10}       
\end{figure*}

A new experimental setup to measure prompt-delayed $\gamma$-ray coincidences from 
isotopically identified fission fragments, over a wide time range, $100$~ns - $200$~$\mu$s, 
is presented in Ref.~\cite{Kim2017}. The fission fragments are isotopically identified, 
on an event-by-event basis, using the VAMOS++ large acceptance spectrometer. 
The prompt $\gamma$ rays ($\gamma_P$) emitted near the target were detected using 
the AGATA $\gamma$-ray tracking array .
The fission fragments reaching the focal plane after a typical time-of-flight
of $\sim 200$~ns, were stopped in the ionisation
chamber. Delayed $\gamma$ rays ($\gamma_D$) were detected using seven EXOGAM
HPGe Clover detectors~\cite{Sim00} arranged in a wall-like
configuration at the focal plane of the VAMOS++
spectrometer. A $2$~mm thick aluminium window between the ionisation chamber and the Clover
detectors was used to minimise the attenuation of the
emitted $\gamma$ rays. A $3$~mm thick lead shielding was placed
after and in-between the Clover detectors to minimise the
events arising from the room-background and Compton
scattering between the Clover detectors.
The details of the experimental setup and analysis methods are discussed in Ref.~\cite{Kim2017}. 

The results obtained using this experimental set-up will be illustrated based on the
case of well studied $^{132}$Te~\cite{Wolf1976, Fogelberg1986, Astier2012, McDonald1973,
Hughes2005, Roberts2013, Genevey2001, Biswas2016}.
Figure~\ref{fig:9}(a) shows the partial level scheme of $^{132}$Te 
(below $4.3$~MeV). Several different isomeric states have been reported for this nucleus, in particular, the $7^-$ excited state at $1925$~keV
with the half-life of $t_{1/2} = 28.1$~$\mu$s~\cite{McDonald1973}.
In earlier works the prompt transitions $1040$~keV and  $292$~keV have been observed and placed tentatively in the level 
scheme as feeding the $7^-$ state~\cite{Hughes2005, Biswas2016}.
However, the prompt-delayed correlation between the $\gamma$ rays
populating and depopulating the $7^-$ state could not be observed.
In Fig.~\ref{fig:9}(b) the Doppler corrected prompt $\gamma$ ray ($\gamma_P$) spectrum
observed in coincidence with delayed $E_{\gamma_D} = 151$~keV $\gamma$-ray, which depopulates 
the $7^-$ isomeric state, is shown. The prompt $\gamma$ rays $292$~keV, $624$~keV, 
$900$~keV and $1040$~keV, are seen in the spectrum. Figure~\ref{fig:9}(c)
shows the delayed $\gamma$-ray spectrum in coincidence with the prompt $E_{\gamma_P}$ at $1040$~keV.
The delayed $\gamma$ rays $103$~keV, $151$~keV, 
$697$~keV and $974$~keV, are seen in the spectrum.
This measurement confirms experimentally proposed the level scheme, and illustrates the capabilities of the VAMOS-AGATA-EXOGAM setup to properly correlate prompt and delayed $\gamma$ rays across a long-lived isomer. Further details are discussed in  Ref.~\cite{Kim2017}.

\subsection{Prompt-delayed $\gamma$-ray spectroscopy of neutron-rich isotopes of Sb}

The $Z=50$ shell closure, near $N=82$, is unique in the sense that it is the only 
shell closure with the spin-orbit partner high-spin orbitals, $\pi$g$_{9/2}$ and $\pi$g$_{7/2}$, 
enclosing the magic gap. The interaction of the proton hole/particle in the above-mentioned 
orbitals with neutrons in the high-spin $\nu$h$_{11/2}$ orbital is an important prerequisite 
to the understanding of the nuclear structure near $N=82$ and establishing the features of 
the $\nu \pi$ interaction. 

\begin{figure}[ht]
\centering
  \includegraphics[width=0.8\columnwidth]{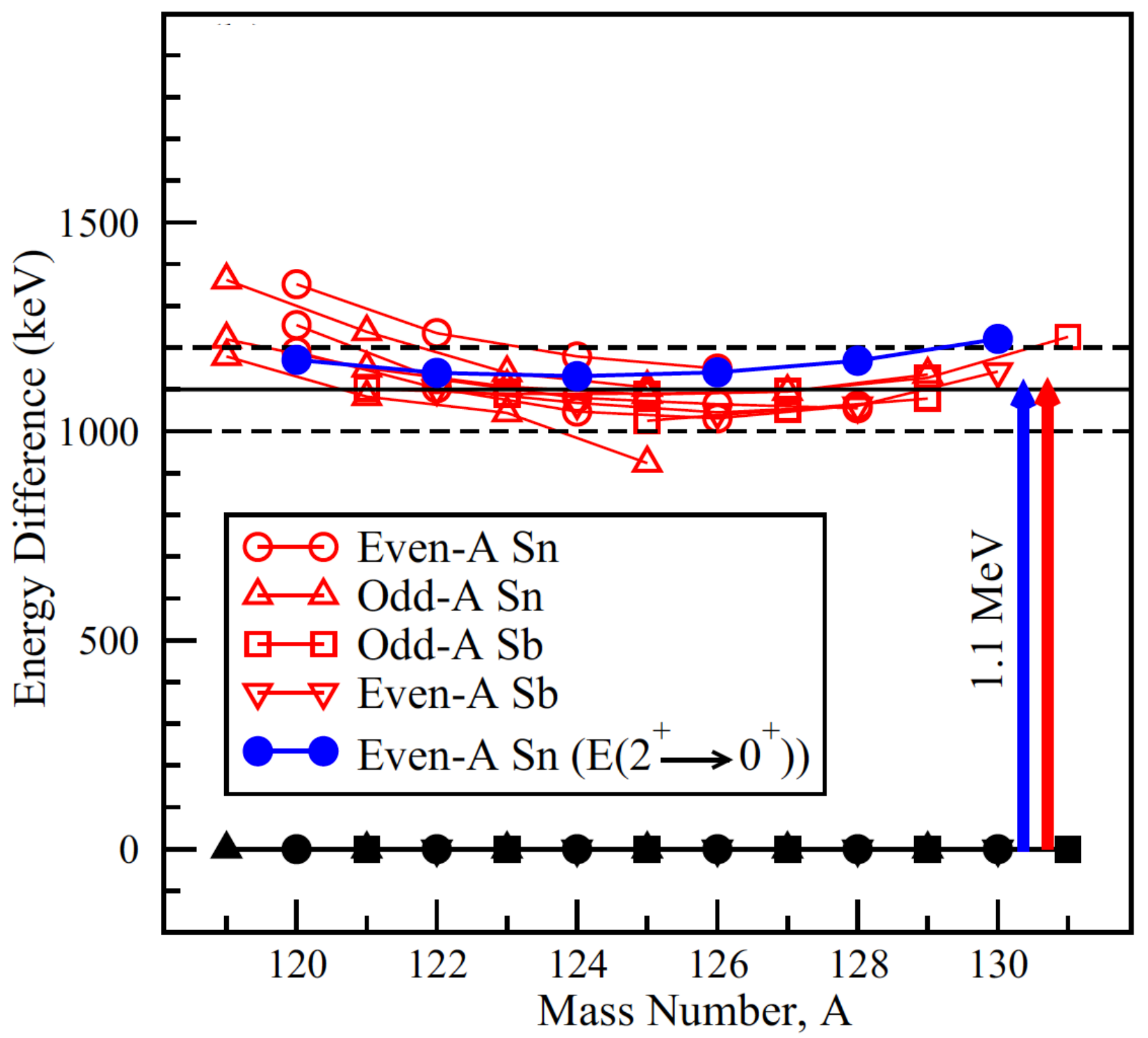}

\caption{
Plot of the energy differences between states with neutron seniority 
$\upsilon_\nu = n$ and $\upsilon_\nu = n + 2 (n = 0, \ldots , 4)$, for odd-$A$
and even-$A$ $^{119 - 130}$Sn and $^{121– 131}$Sb nuclei.  Figure reproduced from \cite{Biswas2019}~\href{https://creativecommons.org/licenses/by-sa/4.0/}{CC BY 4.0}. \copyright~2017, S. Biswas {\it et al.}, published by American Physical Society}
\label{fig:11}
\end{figure}

Prompt-delayed $\gamma$ ray of high-spin states in neutron-rich
$^{122-131}$Sb ($Z=51$)~\cite{Biswas2019},
$^{130-134}$I ($Z=53$)~\cite{Banik2020a} and $^{119,121}$In ($Z=49$)~\cite{Biswas2020},
using the unique experimental setup combining AGATA, VAMOS++ and EXOGAM were reported. This experiment is 
referred to as the Exp.~1, see Table~\ref{tab:ExpSummary}.
In this section we will restrict ourselves to the isotopes of Sb~\cite{Biswas2019}.

Figure~\ref{fig:10} shows the level schemes of $^{125–128}$Sb. 
The newly observed $\gamma$-ray transitions above and below the isomer are indicated in red and blue,
respectively. Previously known half-lives have been remeasured and are underlined 
by a red line, whereas the newly measured half-lives are marked with a red box.
The wealth of the new experimental data obtained in Ref.~\cite{Biswas2019} can be clearly seen from the figure.

The experimental data was compared with theoretical results obtained from LSSM.
A consistent agreement with the excitation energies
and the $B(E2)$ transition probabilities in neutron-rich Sn and Sb isotopes was obtained. 
The isomeric configurations in Sn and Sb were found to be relatively pure. 
The LSSM calculations revealed that the presence of a single valence proton, 
mainly in the $\pi$g$_{7/2}$ orbital in Sb isotopes, leads to significant mixing,
due to the $\nu \pi$ interaction, of
(i) the neutron seniorities ($\upsilon_\nu$)\footnote{$\upsilon_\nu$ stands for neutron seniority,
which refers to the number of unpaired neutrons} and (ii) the neutron angular momentum ($I_\nu$). 
The above features have a weak impact on the excitation energies, but have an important impact 
on the nuclear wave function of the excited states and thus on the corresponding $B(E2)$ 
transition probabilities. 
In addition, a striking feature of the constancy of the energy differences,
where the increase in the number of broken neutron pairs is involved, was observed.
A plot of such energy differences in $^{119–130}$Sn
and $^{122–131}$Sb isotopes is shown in Fig.~\ref{fig:11}. 
This figure shows that the average energy for the
breaking of the first and second pair of neutrons is $\sim 1.1$~MeV,
and that this energy is constant (with a deviation of $\sim100$~keV) for a wide
range of mass numbers, irrespective of the excitation energy
and mixing of neutron seniorities ($\upsilon_\nu$) in the case of Sn and
Sb. In addition, it follows the behavior of even-$A$ Sn isotopes
for $E(2^+ \rightarrow 0^+)$. Further details are discussed in Ref.~\cite{Biswas2019}.

\section{Summary and Conclusions}

Among a large variety of experiments performed at GANIL using the AGATA $\gamma$-ray array,
four have focussed on the nuclear structure studies of isotopically identified fission 
fragments employing the VAMOS++ magnetic spectrometer in coincidence.

The combination of the AGATA $\gamma$-ray array with the VAMOS++ spectrometer forms a unique,
highly performant experimental setup combining efficiency\-
with counting rate capabilities, as well as selectivity with excellent Doppler 
correction of $\gamma$-ray energy and precise isotopic identification.
The performed experiments have been very fruitful and numerous pertinent results have been 
obtained including the $\gamma$-ray spectroscopy of $^{96}$Kr~\cite{Dudouet2017}, $^{81}$Ga~\cite{Dudouet2019}, $^{83,85,87}${As}~\cite{Rezynkina2022}, lifetime
measurements of excited states using the RDDS method in $^{84}$Ge, $^{88}$Kr, 
$^{86}$Se~\cite{Delafosse2018, Delafosse2019},  neutron-rich Zr, Mo and 
Ru~\cite{PhDAnsari} and prompt-delayed $\gamma$-ray 
spectroscopy, using the EXOGAM array for delayed $\gamma$ rays,
of $^{122-131}$Sb~\cite{Biswas2019},
$^{119-121}$In~\cite{Biswas2020} and $^{130-134}$I~\cite{Banik2020a}. 

In the future, fission fragments spectroscopy program will be pursued
at LNL using combination of AGATA and the PRISMA~\cite{Montagnoli2005}
spectrometer. The ongoing development of $^{238}$U beams at energies
around the Coulomb barrier will extend measurements using inverse
kinematics reactions in addition to presently available $^{208}$Pb
beams.  Further, the increased number of available AGATA crystals will
allow to cover 2$\pi$ solid angle at the nominal detector distance,
effectively doubling the solid angle coverage compared to the
experiments presented in this work.  This will improve the
$\gamma-\gamma$ coincidence efficiency, allowing to expand
investigations of exotic neutron-rich nuclei by fission-fragment
spectroscopy.

\section*{Acknowledgments}
The authors thank the AGATA collaboration, the e661, e680, e669 and
e706 GANIL experimental collaborations and the technical teams at Grand
Accélérateur National d’Ions Lourds for their support during the
experiments. A.G. has received funding from the Norwegian Research
Council, project 325714.

\bibliographystyle{myapsrev4-1} 
\bibliography{Biblio.bib}

\begin{thebibliography}{62}%
\makeatletter
\providecommand \@ifxundefined [1]{%
 \@ifx{#1\undefined}
}%
\providecommand \@ifnum [1]{%
 \ifnum #1\expandafter \@firstoftwo
 \else \expandafter \@secondoftwo
 \fi
}%
\providecommand \@ifx [1]{%
 \ifx #1\expandafter \@firstoftwo
 \else \expandafter \@secondoftwo
 \fi
}%
\providecommand \natexlab [1]{#1}%
\providecommand \enquote  [1]{``#1''}%
\providecommand \bibnamefont  [1]{#1}%
\providecommand \bibfnamefont [1]{#1}%
\providecommand \citenamefont [1]{#1}%
\providecommand \href@noop [0]{\@secondoftwo}%
\providecommand \href [0]{\begingroup \@sanitize@url \@href}%
\providecommand \@href[1]{\@@startlink{#1}\@@href}%
\providecommand \@@href[1]{\endgroup#1\@@endlink}%
\providecommand \@sanitize@url [0]{\catcode `\\12\catcode `\$12\catcode
  `\&12\catcode `\#12\catcode `\^12\catcode `\_12\catcode `\%12\relax}%
\providecommand \@@startlink[1]{}%
\providecommand \@@endlink[0]{}%
\providecommand \url  [0]{\begingroup\@sanitize@url \@url }%
\providecommand \@url [1]{\endgroup\@href {#1}{\urlprefix }}%
\providecommand \urlprefix  [0]{URL }%
\providecommand \Eprint [0]{\href }%
\providecommand \doibase [0]{http://dx.doi.org/}%
\providecommand \selectlanguage [0]{\@gobble}%
\providecommand \bibinfo  [0]{\@secondoftwo}%
\providecommand \bibfield  [0]{\@secondoftwo}%
\providecommand \translation [1]{[#1]}%
\providecommand \BibitemOpen [0]{}%
\providecommand \bibitemStop [0]{}%
\providecommand \bibitemNoStop [0]{.\EOS\space}%
\providecommand \EOS [0]{\spacefactor3000\relax}%
\providecommand \BibitemShut  [1]{\csname bibitem#1\endcsname}%
\let\auto@bib@innerbib\@empty
\bibitem [{\citenamefont {Hamilton}\ \emph {et~al.}(1995)\citenamefont
  {Hamilton}, \citenamefont {Ramayya}, \citenamefont {Zhu}, \citenamefont
  {Ter-Akopian}, \citenamefont {Oganessian}, \citenamefont {Cole},
  \citenamefont {Rasmussen},\ and\ \citenamefont {Stoyer}}]{Hamilton1995}%
  \BibitemOpen
  \bibfield  {author} {\bibinfo {author} {\bibfnamefont {J.}~\bibnamefont
  {Hamilton}},  \emph {et~al.},\ }\href {\doibase
  https://doi.org/10.1016/0146-6410(95)00048-N} {\bibfield  {journal} {\bibinfo
   {journal} {Prog. Part. Nucl. Phys.}\ }\textbf {\bibinfo {volume} {35}},\
  \bibinfo {pages} {635} (\bibinfo {year} {1995})}\BibitemShut {NoStop}%
\bibitem [{\citenamefont {Ahmad}(1995)\citenamefont {Ahmad}\ and\ \citenamefont
  {Phillips}}]{Ahmad1995}%
  \BibitemOpen
  \bibfield  {author} {\bibinfo {author} {\bibfnamefont {I.}~\bibnamefont
  {Ahmad}}\ and\ \bibinfo {author} {\bibfnamefont {W.~R.}\ \bibnamefont
  {Phillips}},\ }\href {\doibase 10.1088/0034-4885/58/11/002} {\bibfield
  {journal} {\bibinfo  {journal} {Rep. Prog. Phys.}\ }\textbf {\bibinfo
  {volume} {58}},\ \bibinfo {pages} {1415} (\bibinfo {year}
  {1995})}\BibitemShut {NoStop}%
\bibitem [{\citenamefont {Navin}\ \emph {et~al.}(2014)\citenamefont {Navin},
  \citenamefont {Rejmund}, \citenamefont {Schmitt}, \citenamefont
  {Bhattacharyya}, \citenamefont {Lhersonneau}, \citenamefont {{Van Isacker}},
  \citenamefont {Caama\~{n}o}, \citenamefont {Cl\'{e}ment}, \citenamefont
  {Delaune}, \citenamefont {Farget}, \citenamefont {de~France},\ and\
  \citenamefont {Jacquot}}]{Navin2014}%
  \BibitemOpen
  \bibfield  {author} {\bibinfo {author} {\bibfnamefont {A.}~\bibnamefont
  {Navin}},  \emph {et~al.},\ }\href {\doibase 10.1016/j.physletb.2013.11.024}
  {\bibfield  {journal} {\bibinfo  {journal} {Phys. Lett. B}\ }\textbf
  {\bibinfo {volume} {728}},\ \bibinfo {pages} {136} (\bibinfo {year}
  {2014})}\BibitemShut {NoStop}%
\bibitem [{\citenamefont {Navin}(2014)\citenamefont {Navin}\ and\ \citenamefont
  {Rejmund}}]{Nav2014Yb}%
  \BibitemOpen
  \bibfield  {author} {\bibinfo {author} {\bibfnamefont {A.}~\bibnamefont
  {Navin}}\ and\ \bibinfo {author} {\bibfnamefont {M.}~\bibnamefont
  {Rejmund}},\ }in\ \href {\doibase 10.1036/1097-8542.YB140316} {\emph
  {\bibinfo {booktitle} {McGRAW-HILL Yearbook of Science and Technology}}}\
  (\bibinfo {year} {2014})\ p.\ \bibinfo {pages} {137}\BibitemShut {NoStop}%
\bibitem [{\citenamefont {Leoni}\ \emph {et~al.}(2022)\citenamefont {Leoni},
  \citenamefont {Michelagnoli},\ and\ \citenamefont {Wilson}}]{Leoni2021}%
  \BibitemOpen
  \bibfield  {author} {\bibinfo {author} {\bibfnamefont {S.}~\bibnamefont
  {Leoni}}, \bibinfo {author} {\bibfnamefont {C.}~\bibnamefont {Michelagnoli}},
  \ and\ \bibinfo {author} {\bibfnamefont {J.~N.}\ \bibnamefont {Wilson}},\
  }\href {https://doi.org/10.1007/s40766-022-00033-2} {\bibfield  {journal}
  {\bibinfo  {journal} {La Rivista del Nuovo Cimento}\ }\textbf {\bibinfo
  {volume} {45}},\ \bibinfo {pages} {461} (\bibinfo {year} {2022})}\BibitemShut
  {NoStop}%
\bibitem [{\citenamefont {Metag}\ \emph {et~al.}(1980)\citenamefont {Metag},
  \citenamefont {Habs},\ and\ \citenamefont {Specht}}]{Metag1980}%
  \BibitemOpen
  \bibfield  {author} {\bibinfo {author} {\bibfnamefont {V.}~\bibnamefont
  {Metag}}, \bibinfo {author} {\bibfnamefont {D.}~\bibnamefont {Habs}}, \ and\
  \bibinfo {author} {\bibfnamefont {H.}~\bibnamefont {Specht}},\ }\href
  {\doibase https://doi.org/10.1016/0370-1573(80)90006-X} {\bibfield  {journal}
  {\bibinfo  {journal} {Physics Reports}\ }\textbf {\bibinfo {volume} {65}},\
  \bibinfo {pages} {1} (\bibinfo {year} {1980})}\BibitemShut {NoStop}%
\bibitem [{\citenamefont {Lee}(1997)}]{Lee1997}%
  \BibitemOpen
  \bibfield  {author} {\bibinfo {author} {\bibfnamefont {I.}~\bibnamefont
  {Lee}},\ }\href {\doibase https://doi.org/10.1016/S0146-6410(97)00009-4}
  {\bibfield  {journal} {\bibinfo  {journal} {Prog. Part. Nucl. Phys.}\
  }\textbf {\bibinfo {volume} {38}},\ \bibinfo {pages} {65} (\bibinfo {year}
  {1997})},\ \bibinfo {note} {$4\pi$ High Resolution Gamma Ray Spectroscopy and
  Nuclear Structure}\BibitemShut {NoStop}%
\bibitem [{\citenamefont {Urban}\ \emph {et~al.}(1997)\citenamefont {Urban},
  \citenamefont {Durell}, \citenamefont {Phillips}, \citenamefont {Smith},
  \citenamefont {Jones}, \citenamefont {Ahmad}, \citenamefont {Barnett},
  \citenamefont {Bentaleb}, \citenamefont {Dorning}, \citenamefont {Leddy},
  \citenamefont {Lubkiewicz}, \citenamefont {Morss}, \citenamefont
  {Rzaca-Urban}, \citenamefont {Sareen}, \citenamefont {Schulz},\ and\
  \citenamefont {Varley}}]{Urban1997}%
  \BibitemOpen
  \bibfield  {author} {\bibinfo {author} {\bibfnamefont {W.}~\bibnamefont
  {Urban}},  \emph {et~al.},\ }\href {\doibase 10.1007/s002180050291}
  {\bibfield  {journal} {\bibinfo  {journal} {Z. Phys. A}\ }\textbf {\bibinfo
  {volume} {358}},\ \bibinfo {pages} {145} (\bibinfo {year}
  {1997})}\BibitemShut {NoStop}%
\bibitem [{\citenamefont {Korten}(2003)\citenamefont {Korten}\ and\
  \citenamefont {Lunardi}}]{EUROBALL}%
  \BibitemOpen
  \bibfield  {author} {\bibinfo {author} {\bibfnamefont {W.}~\bibnamefont
  {Korten}}\ and\ \bibinfo {author} {\bibfnamefont {S.}~\bibnamefont
  {Lunardi}},\ }\href {http://euroball.lnl.infn.it/EBmore/EB_Final_Report.pdf}
  {\emph {\bibinfo {title} {Achievements with the Euroball spectrometer
  (1997-2003)}}}\ (\bibinfo {year} {2003})\BibitemShut {NoStop}%
\bibitem [{\citenamefont {Jentschel}\ \emph {et~al.}(2017)\citenamefont
  {Jentschel}, \citenamefont {Blanc}, \citenamefont {de~France}, \citenamefont
  {Köster}, \citenamefont {Leoni}, \citenamefont {Mutti}, \citenamefont
  {Simpson}, \citenamefont {Soldner}, \citenamefont {Ur}, \citenamefont
  {Urban}, \citenamefont {Ahmed}, \citenamefont {Astier}, \citenamefont
  {Augey}, \citenamefont {Back}, \citenamefont {Ba{\c}czyk}, \citenamefont
  {Bajoga}, \citenamefont {Balabanski}, \citenamefont {Belgya}, \citenamefont
  {Benzoni}, \citenamefont {Bernards}, \citenamefont {Biswas}, \citenamefont
  {Bocchi}, \citenamefont {Bottoni}, \citenamefont {Britton}, \citenamefont
  {Bruyneel}, \citenamefont {Burnett}, \citenamefont {Cakirli}, \citenamefont
  {Carroll}, \citenamefont {Catford}, \citenamefont {Cederwall}, \citenamefont
  {Celikovic}, \citenamefont {Cieplicka-Ory{\'{n}}czak}, \citenamefont
  {Clement}, \citenamefont {Cooper}, \citenamefont {Crespi}, \citenamefont
  {Csatlos}, \citenamefont {Curien}, \citenamefont {Czerwi{\'{n}}ski},
  \citenamefont {Danu}, \citenamefont {Davies}, \citenamefont {Didierjean},
  \citenamefont {Drouet}, \citenamefont {Duch{\^{e}}ne}, \citenamefont
  {Ducoin}, \citenamefont {aM. Jentschel}, \citenamefont {Blanc}, \citenamefont
  {de~France}, \citenamefont {Köster}, \citenamefont {Leoni}, \citenamefont
  {Mutti}, \citenamefont {Simpson}, \citenamefont {Soldner}, \citenamefont
  {Ur}, \citenamefont {Urban}, \citenamefont {Ahmed}, \citenamefont {Astier},
  \citenamefont {Augey}, \citenamefont {Back}, \citenamefont {Ba{\c}czyk},
  \citenamefont {Bajoga}, \citenamefont {Balabanski}, \citenamefont {Belgya},
  \citenamefont {Benzoni}, \citenamefont {Bernards}, \citenamefont {Biswas},
  \citenamefont {Bocchi}, \citenamefont {Bottoni}, \citenamefont {Britton},
  \citenamefont {Bruyneel}, \citenamefont {Burnett}, \citenamefont {Cakirli},
  \citenamefont {Carroll}, \citenamefont {Catford}, \citenamefont {Cederwall},
  \citenamefont {Celikovic}, \citenamefont {Cieplicka-Ory{\'{n}}czak},
  \citenamefont {Clement}, \citenamefont {Cooper}, \citenamefont {Crespi},
  \citenamefont {Csatlos}, \citenamefont {Curien}, \citenamefont
  {Czerwi{\'{n}}ski}, \citenamefont {Danu}, \citenamefont {Davies},
  \citenamefont {Didierjean}, \citenamefont {Drouet}, \citenamefont
  {Duch{\^{e}}ne}, \citenamefont {Ducoin}, \citenamefont {Eberhardt},
  \citenamefont {Erturk}, \citenamefont {Fraile}, \citenamefont {Gottardo},
  \citenamefont {Grente}, \citenamefont {Grocutt}, \citenamefont {Guerrero},
  \citenamefont {Guinet}, \citenamefont {Hartig}, \citenamefont {Henrich},
  \citenamefont {Ignatov}, \citenamefont {Ilieva}, \citenamefont {Ivanova},
  \citenamefont {John}, \citenamefont {John}, \citenamefont {Jolie},
  \citenamefont {Kisyov}, \citenamefont {Krticka}, \citenamefont
  {Konstantinopoulos}, \citenamefont {Korgul}, \citenamefont {Krasznahorkay},
  \citenamefont {Kröll}, \citenamefont {Kurpeta}, \citenamefont {Kuti},
  \citenamefont {Lalkovski}, \citenamefont {Larijani}, \citenamefont
  {Leguillon}, \citenamefont {Lica}, \citenamefont {Litaize}, \citenamefont
  {Lozeva}, \citenamefont {Magron}, \citenamefont {Mancuso}, \citenamefont
  {Martinez}, \citenamefont {Massarczyk}, \citenamefont {Mazzocchi},
  \citenamefont {Melon}, \citenamefont {Mengoni}, \citenamefont {Michelagnoli},
  \citenamefont {Million}, \citenamefont {Mokry}, \citenamefont {Mukhopadhyay},
  \citenamefont {Mulholland}, \citenamefont {Nannini}, \citenamefont {Napoli},
  \citenamefont {Olaizola}, \citenamefont {Orlandi}, \citenamefont {Patel},
  \citenamefont {Paziy}, \citenamefont {Petrache}, \citenamefont {Pfeiffer},
  \citenamefont {Pietralla}, \citenamefont {Podolyak}, \citenamefont
  {Ramdhane}, \citenamefont {Redon}, \citenamefont {Regan}, \citenamefont
  {Regis}, \citenamefont {Regnier}, \citenamefont {Oliver}, \citenamefont
  {Rudigier}, \citenamefont {Runke}, \citenamefont {Rza{\c}ca-Urban},
  \citenamefont {Saed-Samii}, \citenamefont {Salsac}, \citenamefont {Scheck},
  \citenamefont {Schwengner}, \citenamefont {Sengele}, \citenamefont {Singh},
  \citenamefont {Smith}, \citenamefont {Stezowski}, \citenamefont {Szpak},
  \citenamefont {Thomas}, \citenamefont {Thürauf}, \citenamefont {Timar},
  \citenamefont {Tom}, \citenamefont {Tomandl}, \citenamefont {Tornyi},
  \citenamefont {Townsley}, \citenamefont {Tuerler}, \citenamefont {Valenta},
  \citenamefont {Vancraeyenest}, \citenamefont {Vandone}, \citenamefont
  {Vanhoy}, \citenamefont {Vedia}, \citenamefont {Warr}, \citenamefont
  {Werner}, \citenamefont {Wilmsen}, \citenamefont {Wilson}, \citenamefont
  {Zerrouki}, \citenamefont {Zielinska}, \citenamefont {Erturk}, \citenamefont
  {Fraile}, \citenamefont {Gottardo}, \citenamefont {Grente}, \citenamefont
  {Grocutt}, \citenamefont {Guerrero}, \citenamefont {Guinet}, \citenamefont
  {Hartig}, \citenamefont {Henrich}, \citenamefont {Ignatov}, \citenamefont
  {Ilieva}, \citenamefont {Ivanova}, \citenamefont {John}, \citenamefont
  {John}, \citenamefont {Jolie}, \citenamefont {Kisyov}, \citenamefont
  {Krticka}, \citenamefont {Konstantinopoulos}, \citenamefont {Korgul},
  \citenamefont {Krasznahorkay}, \citenamefont {Kröll}, \citenamefont
  {Kurpeta}, \citenamefont {Kuti}, \citenamefont {Lalkovski}, \citenamefont
  {Larijani}, \citenamefont {Leguillon}, \citenamefont {Lica}, \citenamefont
  {Litaize}, \citenamefont {Lozeva}, \citenamefont {Magron}, \citenamefont
  {Mancuso}, \citenamefont {Martinez}, \citenamefont {Massarczyk},
  \citenamefont {Mazzocchi}, \citenamefont {Melon}, \citenamefont {Mengoni},
  \citenamefont {Michelagnoli}, \citenamefont {Million}, \citenamefont {Mokry},
  \citenamefont {Mukhopadhyay}, \citenamefont {Mulholland}, \citenamefont
  {Nannini}, \citenamefont {Napoli}, \citenamefont {Olaizola}, \citenamefont
  {Orlandi}, \citenamefont {Patel}, \citenamefont {Paziy}, \citenamefont
  {Petrache}, \citenamefont {Pfeiffer}, \citenamefont {Pietralla},
  \citenamefont {Podolyak}, \citenamefont {Ramdhane}, \citenamefont {Redon},
  \citenamefont {Regan}, \citenamefont {Regis}, \citenamefont {Regnier},
  \citenamefont {Oliver}, \citenamefont {Rudigier}, \citenamefont {Runke},
  \citenamefont {Rza{\c}ca-Urban}, \citenamefont {Saed-Samii}, \citenamefont
  {Salsac}, \citenamefont {Scheck}, \citenamefont {Schwengner}, \citenamefont
  {Sengele}, \citenamefont {Singh}, \citenamefont {Smith}, \citenamefont
  {Stezowski}, \citenamefont {Szpak}, \citenamefont {Thomas}, \citenamefont
  {Thürauf}, \citenamefont {Timar}, \citenamefont {Tom}, \citenamefont
  {Tomandl}, \citenamefont {Tornyi}, \citenamefont {Townsley}, \citenamefont
  {Tuerler}, \citenamefont {Valenta}, \citenamefont {Vancraeyenest},
  \citenamefont {Vandone}, \citenamefont {Vanhoy}, \citenamefont {Vedia},
  \citenamefont {Warr}, \citenamefont {Werner}, \citenamefont {Wilmsen},
  \citenamefont {Wilson}, \citenamefont {Zerrouki},\ and\ \citenamefont
  {Zielinska}}]{Jentschel_2017}%
  \BibitemOpen
  \bibfield  {author} {\bibinfo {author} {\bibfnamefont {M.}~\bibnamefont
  {Jentschel}},  \emph {et~al.},\ }\href {\doibase
  10.1088/1748-0221/12/11/p11003} {\bibfield  {journal} {\bibinfo  {journal}
  {J. Instrum.}\ }\textbf {\bibinfo {volume} {12}},\ \bibinfo {pages} {P11003}
  (\bibinfo {year} {2017})}\BibitemShut {NoStop}%
\bibitem [{\citenamefont {Lebois}\ \emph {et~al.}(2020)\citenamefont {Lebois},
  \citenamefont {Jovančević}, \citenamefont {Thisse}, \citenamefont
  {Canavan}, \citenamefont {Étasse}, \citenamefont {Rudigier},\ and\
  \citenamefont {Wilson}}]{Lebois2020}%
  \BibitemOpen
  \bibfield  {author} {\bibinfo {author} {\bibfnamefont {M.}~\bibnamefont
  {Lebois}},  \emph {et~al.},\ }\href {\doibase
  https://doi.org/10.1016/j.nima.2020.163580} {\bibfield  {journal} {\bibinfo
  {journal} {Nucl. Instrum. Methods Phys. Res. A}\ }\textbf {\bibinfo {volume}
  {960}},\ \bibinfo {pages} {163580} (\bibinfo {year} {2020})}\BibitemShut
  {NoStop}%
\bibitem [{\citenamefont {Kim}\ \emph {et~al.}(2017{\natexlab{a}})\citenamefont
  {Kim}, \citenamefont {Lemasson}, \citenamefont {Rejmund}, \citenamefont
  {Navin}, \citenamefont {Biswas}, \citenamefont {Michelagnoli}, \citenamefont
  {Stefan}, \citenamefont {Banik}, \citenamefont {Bednarczyk}, \citenamefont
  {Bhattacharya}, \citenamefont {Bhattacharyya}, \citenamefont {Cl{\'{e}}ment},
  \citenamefont {Crawford}, \citenamefont {{De France}}, \citenamefont
  {Fallon}, \citenamefont {Goupil}, \citenamefont {Jacquot}, \citenamefont
  {Li}, \citenamefont {Ljungvall}, \citenamefont {Macchiavelli}, \citenamefont
  {Maj}, \citenamefont {M{\'{e}}nager}, \citenamefont {Morel}, \citenamefont
  {Palit}, \citenamefont {P{\'{e}}rez-Vidal}, \citenamefont {Ropert},\ and\
  \citenamefont {Schmitt}}]{Kim2017}%
  \BibitemOpen
  \bibfield  {author} {\bibinfo {author} {\bibfnamefont {Y.}~\bibnamefont
  {Kim}},  \emph {et~al.},\ }\href {\doibase 10.1140/epja/i2017-12353-y}
  {\bibfield  {journal} {\bibinfo  {journal} {Eur. Phys. J. A}\ }\textbf
  {\bibinfo {volume} {53}},\ \bibinfo {pages} {162} (\bibinfo {year}
  {2017}{\natexlab{a}})}\BibitemShut {NoStop}%
\bibitem [{\citenamefont {Biswas}\ \emph {et~al.}(2019)\citenamefont {Biswas},
  \citenamefont {Lemasson}, \citenamefont {Rejmund}, \citenamefont {Navin},
  \citenamefont {Kim}, \citenamefont {Michelagnoli}, \citenamefont {Stefan},
  \citenamefont {Banik}, \citenamefont {Bednarczyk}, \citenamefont
  {Bhattacharya}, \citenamefont {Bhattacharyya}, \citenamefont {Cl{\'{e}}ment},
  \citenamefont {Crawford}, \citenamefont {de~France}, \citenamefont {Fallon},
  \citenamefont {Fr{\'{e}}mont}, \citenamefont {Goupil}, \citenamefont
  {Jacquot}, \citenamefont {Li}, \citenamefont {Ljungvall}, \citenamefont
  {Maj}, \citenamefont {M{\'{e}}nager}, \citenamefont {Morel}, \citenamefont
  {Palit}, \citenamefont {P{\'{e}}rez-Vidal}, \citenamefont {Ropert},
  \citenamefont {Barrientos}, \citenamefont {Benzoni}, \citenamefont
  {Birkenbach}, \citenamefont {Boston}, \citenamefont {Boston}, \citenamefont
  {Cederwall}, \citenamefont {Collado}, \citenamefont {Cullen}, \citenamefont
  {D{\'{e}}sesquelles}, \citenamefont {Domingo-Pardo}, \citenamefont {Dudouet},
  \citenamefont {Eberth}, \citenamefont {Gonz{\'{a}}lez}, \citenamefont
  {Harkness-Brennan}, \citenamefont {Hess}, \citenamefont {Jungclaus},
  \citenamefont {Korten}, \citenamefont {Labiche}, \citenamefont {Lefevre},
  \citenamefont {Menegazzo}, \citenamefont {Mengoni}, \citenamefont {Million},
  \citenamefont {Napoli}, \citenamefont {Pullia}, \citenamefont {Quintana},
  \citenamefont {Ralet}, \citenamefont {Recchia}, \citenamefont {Reiter},
  \citenamefont {Saillant}, \citenamefont {Salsac}, \citenamefont {Sanchis},
  \citenamefont {Stezowski}, \citenamefont {Theisen}, \citenamefont
  {Valiente-Dob{\'{o}}n},\ and\ \citenamefont {Zieli{\'{n}}ska}}]{Biswas2019}%
  \BibitemOpen
  \bibfield  {author} {\bibinfo {author} {\bibfnamefont {S.}~\bibnamefont
  {Biswas}},  \emph {et~al.},\ }\href {\doibase 10.1103/PhysRevC.99.064302}
  {\bibfield  {journal} {\bibinfo  {journal} {Phys. Rev. C}\ }\textbf {\bibinfo
  {volume} {99}},\ \bibinfo {pages} {064302} (\bibinfo {year}
  {2019})}\BibitemShut {NoStop}%
\bibitem [{\citenamefont {Biswas}\ \emph {et~al.}(2020)\citenamefont {Biswas},
  \citenamefont {Lemasson}, \citenamefont {Rejmund}, \citenamefont {Navin},
  \citenamefont {Kim}, \citenamefont {Michelagnoli}, \citenamefont {Stefan},
  \citenamefont {Banik}, \citenamefont {Bednarczyk}, \citenamefont
  {Bhattacharya}, \citenamefont {Bhattacharyya}, \citenamefont {Cl{\'{e}}ment},
  \citenamefont {Crawford}, \citenamefont {de~France}, \citenamefont {Fallon},
  \citenamefont {Fr{\'{e}}mont}, \citenamefont {Goupil}, \citenamefont
  {Jacquot}, \citenamefont {Li}, \citenamefont {Ljungvall}, \citenamefont
  {Maj}, \citenamefont {M{\'{e}}nager}, \citenamefont {Morel}, \citenamefont
  {Palit}, \citenamefont {P{\'{e}}rez-Vidal},\ and\ \citenamefont
  {Ropert}}]{Biswas2020}%
  \BibitemOpen
  \bibfield  {author} {\bibinfo {author} {\bibfnamefont {S.}~\bibnamefont
  {Biswas}},  \emph {et~al.},\ }\href {\doibase 10.1103/PhysRevC.102.014326}
  {\bibfield  {journal} {\bibinfo  {journal} {Phys. Rev. C}\ }\textbf {\bibinfo
  {volume} {102}},\ \bibinfo {pages} {014326} (\bibinfo {year}
  {2020})}\BibitemShut {NoStop}%
\bibitem [{\citenamefont {Banik}\ \emph {et~al.}(2020)\citenamefont {Banik},
  \citenamefont {Bhattacharyya}, \citenamefont {Rejmund}, \citenamefont
  {Lemasson}, \citenamefont {Biswas}, \citenamefont {Navin}, \citenamefont
  {Kim}, \citenamefont {Michelagnoli}, \citenamefont {Stefan}, \citenamefont
  {Bednarczyk}, \citenamefont {Bhattacharya}, \citenamefont {Cl{\'{e}}ment},
  \citenamefont {Crawford}, \citenamefont {de~France}, \citenamefont {Fallon},
  \citenamefont {Fr{\'{e}}mont}, \citenamefont {Goupil}, \citenamefont
  {Jacquot}, \citenamefont {Li}, \citenamefont {Ljungvall}, \citenamefont
  {Maj}, \citenamefont {M{\'{e}}nager}, \citenamefont {Morel}, \citenamefont
  {Mukherjee}, \citenamefont {Palit}, \citenamefont {P{\'{e}}rez-Vidal},
  \citenamefont {Ropert},\ and\ \citenamefont {Schmitt}}]{Banik2020a}%
  \BibitemOpen
  \bibfield  {author} {\bibinfo {author} {\bibfnamefont {R.}~\bibnamefont
  {Banik}},  \emph {et~al.},\ }\href {\doibase 10.1103/PhysRevC.102.044329}
  {\bibfield  {journal} {\bibinfo  {journal} {Phys. Rev. C}\ }\textbf {\bibinfo
  {volume} {102}},\ \bibinfo {pages} {044329} (\bibinfo {year}
  {2020})}\BibitemShut {NoStop}%
\bibitem [{\citenamefont {Dudouet}\ \emph {et~al.}(2017)\citenamefont
  {Dudouet}, \citenamefont {Lemasson}, \citenamefont {Duch\^ene}, \citenamefont
  {Rejmund}, \citenamefont {Cl\'ement}, \citenamefont {Michelagnoli},
  \citenamefont {Didierjean}, \citenamefont {Korichi}, \citenamefont {Maquart},
  \citenamefont {Stezowski}, \citenamefont {Lizarazo}, \citenamefont
  {P\'erez-Vidal}, \citenamefont {Andreoiu}, \citenamefont {de~Angelis},
  \citenamefont {Astier}, \citenamefont {Delafosse}, \citenamefont {Deloncle},
  \citenamefont {Dombradi}, \citenamefont {de~France}, \citenamefont {Gadea},
  \citenamefont {Gottardo}, \citenamefont {Jacquot}, \citenamefont {Jones},
  \citenamefont {Konstantinopoulos}, \citenamefont {Kuti}, \citenamefont
  {Le~Blanc}, \citenamefont {Lenzi}, \citenamefont {Li}, \citenamefont
  {Lozeva}, \citenamefont {Million}, \citenamefont {Napoli}, \citenamefont
  {Navin}, \citenamefont {Petrache}, \citenamefont {Pietralla}, \citenamefont
  {Ralet}, \citenamefont {Ramdhane}, \citenamefont {Redon}, \citenamefont
  {Schmitt}, \citenamefont {Sohler}, \citenamefont {Verney}, \citenamefont
  {Barrientos}, \citenamefont {Birkenbach}, \citenamefont {Burrows},
  \citenamefont {Charles}, \citenamefont {Collado}, \citenamefont {Cullen},
  \citenamefont {D\'esesquelles}, \citenamefont {Domingo~Pardo}, \citenamefont
  {Gonz\'alez}, \citenamefont {Harkness-Brennan}, \citenamefont {Hess},
  \citenamefont {Judson}, \citenamefont {Karolak}, \citenamefont {Korten},
  \citenamefont {Labiche}, \citenamefont {Ljungvall}, \citenamefont
  {Menegazzo}, \citenamefont {Mengoni}, \citenamefont {Pullia}, \citenamefont
  {Recchia}, \citenamefont {Reiter}, \citenamefont {Salsac}, \citenamefont
  {Sanchis}, \citenamefont {Theisen}, \citenamefont {Valiente-Dob\'on},\ and\
  \citenamefont {Zieli\ifmmode~\acute{n}\else \'{n}\fi{}ska}}]{Dudouet2017}%
  \BibitemOpen
  \bibfield  {author} {\bibinfo {author} {\bibfnamefont {J.}~\bibnamefont
  {Dudouet}},  \emph {et~al.},\ }\href {\doibase
  10.1103/PhysRevLett.118.162501} {\bibfield  {journal} {\bibinfo  {journal}
  {Phys. Rev. Lett.}\ }\textbf {\bibinfo {volume} {118}},\ \bibinfo {pages}
  {162501} (\bibinfo {year} {2017})}\BibitemShut {NoStop}%
\bibitem [{\citenamefont {Dudouet}\ \emph {et~al.}(2019)\citenamefont
  {Dudouet}, \citenamefont {Lemasson}, \citenamefont {Maquart}, \citenamefont
  {Nowacki}, \citenamefont {Verney}, \citenamefont {Rejmund}, \citenamefont
  {Duch\^ene}, \citenamefont {Stezowski}, \citenamefont {Cl\'ement},
  \citenamefont {Michelagnoli}, \citenamefont {Korichi}, \citenamefont
  {Andreoiu}, \citenamefont {Astier}, \citenamefont {de~Angelis}, \citenamefont
  {de~France}, \citenamefont {Delafosse}, \citenamefont {Deloncle},
  \citenamefont {Didierjean}, \citenamefont {Dombradi}, \citenamefont {Ducoin},
  \citenamefont {Gadea}, \citenamefont {Gottardo}, \citenamefont {Guinet},
  \citenamefont {Jacquot}, \citenamefont {Jones}, \citenamefont
  {Konstantinopoulos}, \citenamefont {Kuti}, \citenamefont {Le~Blanc},
  \citenamefont {Lenzi}, \citenamefont {Li}, \citenamefont {Lozeva},
  \citenamefont {Million}, \citenamefont {Napoli}, \citenamefont {Navin},
  \citenamefont {P\'erez-Vidal}, \citenamefont {Petrache}, \citenamefont
  {Ralet}, \citenamefont {Ramdhane}, \citenamefont {Redon}, \citenamefont
  {Schmitt},\ and\ \citenamefont {Sohler}}]{Dudouet2019}%
  \BibitemOpen
  \bibfield  {author} {\bibinfo {author} {\bibfnamefont {J.}~\bibnamefont
  {Dudouet}},  \emph {et~al.},\ }\href {\doibase 10.1103/PhysRevC.100.011301}
  {\bibfield  {journal} {\bibinfo  {journal} {Phys. Rev. C}\ }\textbf {\bibinfo
  {volume} {100}},\ \bibinfo {pages} {011301} (\bibinfo {year}
  {2019})}\BibitemShut {NoStop}%
\bibitem [{\citenamefont {Rezynkina}\ \emph {et~al.}(2022)\citenamefont
  {Rezynkina}, \citenamefont {Dao}, \citenamefont {Duch\^ene}, \citenamefont
  {Dudouet}, \citenamefont {Nowacki}, \citenamefont {Cl\'ement}, \citenamefont
  {Lemasson}, \citenamefont {Andreoiu}, \citenamefont {Astier}, \citenamefont
  {de~Angelis}, \citenamefont {de~France}, \citenamefont {Delafosse},
  \citenamefont {Deloncle}, \citenamefont {Didierjean}, \citenamefont
  {Dombradi}, \citenamefont {Ducoin}, \citenamefont {Gadea}, \citenamefont
  {Gottardo}, \citenamefont {Guinet}, \citenamefont {Jacquot}, \citenamefont
  {Jones}, \citenamefont {Konstantinopoulos}, \citenamefont {Kuti},
  \citenamefont {Korichi}, \citenamefont {Lenzi}, \citenamefont {Li},
  \citenamefont {Le~Blanc}, \citenamefont {Lizarazo}, \citenamefont {Lozeva},
  \citenamefont {Maquart}, \citenamefont {Million}, \citenamefont
  {Michelagnoli}, \citenamefont {Napoli}, \citenamefont {Navin}, \citenamefont
  {P\'erez-Vidal}, \citenamefont {Petrache}, \citenamefont {Pietralla},
  \citenamefont {Ralet}, \citenamefont {Ramdhane}, \citenamefont {Rejmund},
  \citenamefont {Stezowski}, \citenamefont {Schmitt}, \citenamefont {Sohler},\
  and\ \citenamefont {Verney}}]{Rezynkina2022}%
  \BibitemOpen
  \bibfield  {author} {\bibinfo {author} {\bibfnamefont {K.}~\bibnamefont
  {Rezynkina}},  \emph {et~al.},\ }\href {\doibase 10.1103/PhysRevC.106.014320}
  {\bibfield  {journal} {\bibinfo  {journal} {Phys. Rev. C}\ }\textbf {\bibinfo
  {volume} {106}},\ \bibinfo {pages} {014320} (\bibinfo {year}
  {2022})}\BibitemShut {NoStop}%
\bibitem [{\citenamefont {Delafosse}\ \emph {et~al.}(2018)\citenamefont
  {Delafosse}, \citenamefont {Verney}, \citenamefont
  {Marevi\ifmmode~\acute{c}\else \'{c}\fi{}}, \citenamefont {Gottardo},
  \citenamefont {Michelagnoli}, \citenamefont {Lemasson}, \citenamefont
  {Goasduff}, \citenamefont {Ljungvall}, \citenamefont {Cl\'ement},
  \citenamefont {Korichi}, \citenamefont {De~Angelis}, \citenamefont
  {Andreoiu}, \citenamefont {Babo}, \citenamefont {Boso}, \citenamefont
  {Didierjean}, \citenamefont {Dudouet}, \citenamefont {Franchoo},
  \citenamefont {Gadea}, \citenamefont {Georgiev}, \citenamefont {Ibrahim},
  \citenamefont {Jacquot}, \citenamefont {Konstantinopoulos}, \citenamefont
  {Lenzi}, \citenamefont {Maquart}, \citenamefont {Matea}, \citenamefont
  {Mengoni}, \citenamefont {Napoli}, \citenamefont {Nik\ifmmode \check{s}\else
  \v{s}\fi{}i\ifmmode~\acute{c}\else \'{c}\fi{}}, \citenamefont {Olivier},
  \citenamefont {P\'erez-Vidal}, \citenamefont {Portail}, \citenamefont
  {Recchia}, \citenamefont {Redon}, \citenamefont {Siciliano}, \citenamefont
  {Stefan}, \citenamefont {Stezowski}, \citenamefont {Vretenar}, \citenamefont
  {Zielinska}, \citenamefont {Barrientos}, \citenamefont {Benzoni},
  \citenamefont {Birkenbach}, \citenamefont {Boston}, \citenamefont {Boston},
  \citenamefont {Cederwall}, \citenamefont {Charles}, \citenamefont {Ciemala},
  \citenamefont {Collado}, \citenamefont {Cullen}, \citenamefont
  {D\'esesquelles}, \citenamefont {de~France}, \citenamefont {Domingo-Pardo},
  \citenamefont {Eberth}, \citenamefont {Gonz\'alez}, \citenamefont
  {Harkness-Brennan}, \citenamefont {Hess}, \citenamefont {Judson},
  \citenamefont {Jungclaus}, \citenamefont {Korten}, \citenamefont {Lefevre},
  \citenamefont {Legruel}, \citenamefont {Menegazzo}, \citenamefont {Million},
  \citenamefont {Nyberg}, \citenamefont {Quintana}, \citenamefont {Ralet},
  \citenamefont {Reiter}, \citenamefont {Saillant}, \citenamefont {Sanchis},
  \citenamefont {Theisen},\ and\ \citenamefont
  {Valiente~Dobon}}]{Delafosse2018}%
  \BibitemOpen
  \bibfield  {author} {\bibinfo {author} {\bibfnamefont {C.}~\bibnamefont
  {Delafosse}},  \emph {et~al.},\ }\href {\doibase
  10.1103/PhysRevLett.121.192502} {\bibfield  {journal} {\bibinfo  {journal}
  {Phys. Rev. Lett.}\ }\textbf {\bibinfo {volume} {121}},\ \bibinfo {pages}
  {192502} (\bibinfo {year} {2018})}\BibitemShut {NoStop}%
\bibitem [{\citenamefont {Delafosse}(2019)\citenamefont {Delafosse} \emph
  {et~al.}}]{Delafosse2019}%
  \BibitemOpen
  \bibfield  {author} {\bibinfo {author} {\bibfnamefont {C.}~\bibnamefont
  {Delafosse}} \emph {et~al.},\ }\href {\doibase 10.5506/APhysPolB.50.633}
  {\bibfield  {journal} {\bibinfo  {journal} {Acta Phys. Pol. B}\ }\textbf
  {\bibinfo {volume} {50}},\ \bibinfo {pages} {633} (\bibinfo {year}
  {2019})}\BibitemShut {NoStop}%
\bibitem [{\citenamefont {Ansari}(2019)}]{PhDAnsari}%
  \BibitemOpen
  \bibfield  {author} {\bibinfo {author} {\bibfnamefont {S.}~\bibnamefont
  {Ansari}},\ }\emph {\bibinfo {title} {{Shape evolution in neutron-rich Zr, Mo
  and Ru isotopes around mass A=100}}},\ \href
  {https://tel.archives-ouvertes.fr/tel-02445759} {\bibinfo {type} {Theses}},\
  \bibinfo  {school} {{Universit{\'e} Paris Saclay (COmUE)}} (\bibinfo {year}
  {2019})\BibitemShut {NoStop}%
\bibitem [{\citenamefont {Rejmund}\ \emph {et~al.}(2011)\citenamefont
  {Rejmund}, \citenamefont {Lecornu}, \citenamefont {Navin}, \citenamefont
  {Schmitt}, \citenamefont {Damoy}, \citenamefont {Delaune}, \citenamefont
  {Enguerrand}, \citenamefont {Fremont}, \citenamefont {Gangnant},
  \citenamefont {Gaudefroy}, \citenamefont {Jacquot}, \citenamefont {Pancin},
  \citenamefont {Pullanhiotan},\ and\ \citenamefont {Spitaels}}]{Rejmund2011}%
  \BibitemOpen
  \bibfield  {author} {\bibinfo {author} {\bibfnamefont {M.}~\bibnamefont
  {Rejmund}},  \emph {et~al.},\ }\href {\doibase 10.1016/j.nima.2011.05.007}
  {\bibfield  {journal} {\bibinfo  {journal} {Nucl. Instrum. Methods Phys. Res.
  A}\ }\textbf {\bibinfo {volume} {646}},\ \bibinfo {pages} {184} (\bibinfo
  {year} {2011})}\BibitemShut {NoStop}%
\bibitem [{\citenamefont {Montagnoli}\ \emph {et~al.}(2005)\citenamefont
  {Montagnoli}, \citenamefont {Stefanini}, \citenamefont {Trotta},
  \citenamefont {Beghini}, \citenamefont {Bettini}, \citenamefont
  {Scarlassara}, \citenamefont {Schiavon}, \citenamefont {Corradi},
  \citenamefont {Behera}, \citenamefont {Fioretto}, \citenamefont {Gadea},
  \citenamefont {Latina}, \citenamefont {Szilner}, \citenamefont {Donà},
  \citenamefont {Rigato}, \citenamefont {Kondratiev}, \citenamefont {Chizhov},
  \citenamefont {Kniajeva}, \citenamefont {Kozulin}, \citenamefont
  {Pokrovskiy}, \citenamefont {Voskressensky},\ and\ \citenamefont
  {Ackermann}}]{Montagnoli2005}%
  \BibitemOpen
  \bibfield  {author} {\bibinfo {author} {\bibfnamefont {G.}~\bibnamefont
  {Montagnoli}},  \emph {et~al.},\ }\href {\doibase
  https://doi.org/10.1016/j.nima.2005.03.158} {\bibfield  {journal} {\bibinfo
  {journal} {Nucl. Instrum. Methods Phys. Res. A}\ }\textbf {\bibinfo {volume}
  {547}},\ \bibinfo {pages} {455} (\bibinfo {year} {2005})}\BibitemShut
  {NoStop}%
\bibitem [{\citenamefont {Rejmund}\ \emph
  {et~al.}(2016{\natexlab{a}})\citenamefont {Rejmund}, \citenamefont {Navin},
  \citenamefont {Biswas}, \citenamefont {Lemasson}, \citenamefont
  {Caama{\~{n}}o}, \citenamefont {Cl{\'{e}}ment}, \citenamefont {Delaune},
  \citenamefont {Farget}, \citenamefont {de~France}, \citenamefont {Jacquot},\
  and\ \citenamefont {{Van Isacker}}}]{Rejmund2015a}%
  \BibitemOpen
  \bibfield  {author} {\bibinfo {author} {\bibfnamefont {M.}~\bibnamefont
  {Rejmund}},  \emph {et~al.},\ }\href {\doibase
  10.1016/j.physletb.2015.11.077} {\bibfield  {journal} {\bibinfo  {journal}
  {Phys. Lett. B}\ }\textbf {\bibinfo {volume} {753}},\ \bibinfo {pages} {86}
  (\bibinfo {year} {2016}{\natexlab{a}})}\BibitemShut {NoStop}%
\bibitem [{\citenamefont {Rejmund}\ \emph
  {et~al.}(2016{\natexlab{b}})\citenamefont {Rejmund}, \citenamefont {Navin},
  \citenamefont {Bhattacharyya}, \citenamefont {Caama{\~{n}}o}, \citenamefont
  {Cl{\'{e}}ment}, \citenamefont {Delaune}, \citenamefont {Farget},
  \citenamefont {de~France}, \citenamefont {Jacquot},\ and\ \citenamefont
  {Lemasson}}]{Rejmund2016}%
  \BibitemOpen
  \bibfield  {author} {\bibinfo {author} {\bibfnamefont {M.}~\bibnamefont
  {Rejmund}},  \emph {et~al.},\ }\href {\doibase 10.1103/PhysRevC.93.024312}
  {\bibfield  {journal} {\bibinfo  {journal} {Phys. Rev. C}\ }\textbf {\bibinfo
  {volume} {93}},\ \bibinfo {pages} {024312} (\bibinfo {year}
  {2016}{\natexlab{b}})}\BibitemShut {NoStop}%
\bibitem [{\citenamefont {Simpson}\ \emph {et~al.}(2000)\citenamefont
  {Simpson}, \citenamefont {Azaiez}, \citenamefont {deFrance}, \citenamefont
  {Fouan}, \citenamefont {Gerl}, \citenamefont {Julin}, \citenamefont {Korten},
  \citenamefont {Nolan}, \citenamefont {Nyak\'{o}}, \citenamefont {Sletten},
  \citenamefont {Walker},\ and\ \citenamefont {the
  EXOGAM~Collaboration}}]{Sim00}%
  \BibitemOpen
  \bibfield  {author} {\bibinfo {author} {\bibfnamefont {J.}~\bibnamefont
  {Simpson}},  \emph {et~al.},\ }\href {\doibase 1219-7580} {\bibfield
  {journal} {\bibinfo  {journal} {Acta Phys. Hung. NS-H}\ }\textbf {\bibinfo
  {volume} {11}},\ \bibinfo {pages} {159 } (\bibinfo {year}
  {2000})}\BibitemShut {NoStop}%
\bibitem [{\citenamefont {Kim}\ \emph {et~al.}(2017{\natexlab{b}})\citenamefont
  {Kim}, \citenamefont {Biswas}, \citenamefont {Rejmund}, \citenamefont
  {Navin}, \citenamefont {Lemasson}, \citenamefont {Bhattacharyya},
  \citenamefont {Caama{\~{n}}o}, \citenamefont {Cl{\'{e}}ment}, \citenamefont
  {de~France},\ and\ \citenamefont {Jacquot}}]{Kim2017a}%
  \BibitemOpen
  \bibfield  {author} {\bibinfo {author} {\bibfnamefont {Y.}~\bibnamefont
  {Kim}},  \emph {et~al.},\ }\href {\doibase 10.1016/j.physletb.2017.06.058}
  {\bibfield  {journal} {\bibinfo  {journal} {Phys. Lett. B}\ }\textbf
  {\bibinfo {volume} {772}},\ \bibinfo {pages} {403} (\bibinfo {year}
  {2017}{\natexlab{b}})}\BibitemShut {NoStop}%
\bibitem [{\citenamefont {Navin}\ \emph {et~al.}(2017)\citenamefont {Navin},
  \citenamefont {Rejmund}, \citenamefont {Bhattacharyya}, \citenamefont
  {Palit}, \citenamefont {Bhat}, \citenamefont {Sheikh}, \citenamefont
  {Lemasson}, \citenamefont {Bhattacharya}, \citenamefont {Caama{\~{n}}o},
  \citenamefont {Cl{\'{e}}ment}, \citenamefont {Delaune}, \citenamefont
  {Farget}, \citenamefont {de~France},\ and\ \citenamefont
  {Jacquot}}]{Nav2016}%
  \BibitemOpen
  \bibfield  {author} {\bibinfo {author} {\bibfnamefont {A.}~\bibnamefont
  {Navin}},  \emph {et~al.},\ }\href {\doibase 10.1016/j.physletb.2016.11.020}
  {\bibfield  {journal} {\bibinfo  {journal} {Phys. Lett. B}\ }\textbf
  {\bibinfo {volume} {767}},\ \bibinfo {pages} {480} (\bibinfo {year}
  {2017})}\BibitemShut {NoStop}%
\bibitem [{\citenamefont {Wang}\ \emph {et~al.}(2021)\citenamefont {Wang},
  \citenamefont {Hamilton}, \citenamefont {Ramayya}, \citenamefont {Zachary},
  \citenamefont {Lemasson}, \citenamefont {Navin}, \citenamefont {Rejmund},
  \citenamefont {Bhattacharyya}, \citenamefont {Chen}, \citenamefont {Zhang},
  \citenamefont {Eldridge}, \citenamefont {Hwang}, \citenamefont {Brewer},
  \citenamefont {Luo}, \citenamefont {Rasmussen}, \citenamefont {Zhu},
  \citenamefont {Oganessian}, \citenamefont {Caama{\~{n}}o}, \citenamefont
  {Cl{\'{e}}ment}, \citenamefont {Delaune}, \citenamefont {Farget},
  \citenamefont {France}, \citenamefont {Jacquot}, \citenamefont {Inp},
  \citenamefont {Becquerel},\ and\ \citenamefont {Caen}}]{Wang2021}%
  \BibitemOpen
  \bibfield  {author} {\bibinfo {author} {\bibfnamefont {E.~H.}\ \bibnamefont
  {Wang}},  \emph {et~al.},\ }\href {\doibase 10.1103/PhysRevC.103.034301}
  {\bibfield  {journal} {\bibinfo  {journal} {Phys. Rev. C}\ }\textbf {\bibinfo
  {volume} {103}},\ \bibinfo {pages} {034301} (\bibinfo {year}
  {2021})}\BibitemShut {NoStop}%
\bibitem [{\citenamefont {Wang}\ \emph {et~al.}(2015)\citenamefont {Wang},
  \citenamefont {Lemasson}, \citenamefont {Hamilton}, \citenamefont {Ramayya},
  \citenamefont {Hwang}, \citenamefont {Eldridge}, \citenamefont {Navin},
  \citenamefont {Rejmund}, \citenamefont {Bhattacharyya}, \citenamefont {Liu},
  \citenamefont {Brewer}, \citenamefont {Luo}, \citenamefont {Rasmussen},
  \citenamefont {Liu}, \citenamefont {Zhou}, \citenamefont {Liu}, \citenamefont
  {Li}, \citenamefont {Sun}, \citenamefont {Xu}, \citenamefont {Zhu},
  \citenamefont {Ter-Akopian}, \citenamefont {Oganessian}, \citenamefont
  {Caama{\~{n}}o}, \citenamefont {Cl{\'{e}}ment}, \citenamefont {Delaune},
  \citenamefont {Farget}, \citenamefont {de~France},\ and\ \citenamefont
  {Jacquot}}]{Wang2015}%
  \BibitemOpen
  \bibfield  {author} {\bibinfo {author} {\bibfnamefont {E.~H.}\ \bibnamefont
  {Wang}},  \emph {et~al.},\ }\href {\doibase 10.1103/PhysRevC.92.034317}
  {\bibfield  {journal} {\bibinfo  {journal} {Phys. Rev. C}\ }\textbf {\bibinfo
  {volume} {92}},\ \bibinfo {pages} {034317} (\bibinfo {year}
  {2015})}\BibitemShut {NoStop}%
\bibitem [{\citenamefont {Bhattacharyya}\ \emph {et~al.}(2018)\citenamefont
  {Bhattacharyya}, \citenamefont {Wang}, \citenamefont {Navin}, \citenamefont
  {Rejmund}, \citenamefont {Hamilton}, \citenamefont {Ramayya}, \citenamefont
  {Hwang}, \citenamefont {Lemasson}, \citenamefont {Afanasjev}, \citenamefont
  {Bhattacharya}, \citenamefont {Ranger}, \citenamefont {Caama{\~{n}}o},
  \citenamefont {Cl{\'{e}}ment}, \citenamefont {Delaune}, \citenamefont
  {Farget}, \citenamefont {de~France}, \citenamefont {Jacquot}, \citenamefont
  {Luo}, \citenamefont {Oganessian}, \citenamefont {Rasmussen}, \citenamefont
  {Ter-Akopian},\ and\ \citenamefont {Zhu}}]{Bhattacharyya2018}%
  \BibitemOpen
  \bibfield  {author} {\bibinfo {author} {\bibfnamefont {S.}~\bibnamefont
  {Bhattacharyya}},  \emph {et~al.},\ }\href {\doibase
  10.1103/PhysRevC.98.044316} {\bibfield  {journal} {\bibinfo  {journal} {Phys.
  Rev. C}\ }\textbf {\bibinfo {volume} {98}},\ \bibinfo {pages} {044316}
  (\bibinfo {year} {2018})}\BibitemShut {NoStop}%
\bibitem [{\citenamefont {Dewald}\ \emph {et~al.}(2012)\citenamefont {Dewald},
  \citenamefont {Möller},\ and\ \citenamefont {Petkov}}]{Dewald2012}%
  \BibitemOpen
  \bibfield  {author} {\bibinfo {author} {\bibfnamefont {A.}~\bibnamefont
  {Dewald}}, \bibinfo {author} {\bibfnamefont {O.}~\bibnamefont {Möller}}, \
  and\ \bibinfo {author} {\bibfnamefont {P.}~\bibnamefont {Petkov}},\ }\href
  {\doibase https://doi.org/10.1016/j.ppnp.2012.03.003} {\bibfield  {journal}
  {\bibinfo  {journal} {Prog. Part. Nucl. Phys.}\ }\textbf {\bibinfo {volume}
  {67}},\ \bibinfo {pages} {786} (\bibinfo {year} {2012})}\BibitemShut
  {NoStop}%
\bibitem [{\citenamefont {Singh}\ \emph {et~al.}(2018)\citenamefont {Singh},
  \citenamefont {Korten}, \citenamefont {Hagen}, \citenamefont {G\"orgen},
  \citenamefont {Grente}, \citenamefont {Salsac}, \citenamefont {Farget},
  \citenamefont {Cl\'ement}, \citenamefont {de~France}, \citenamefont
  {Braunroth}, \citenamefont {Bruyneel}, \citenamefont {Celikovic},
  \citenamefont {Delaune}, \citenamefont {Dewald}, \citenamefont {Dijon},
  \citenamefont {Delaroche}, \citenamefont {Girod}, \citenamefont {Hackstein},
  \citenamefont {Jacquot}, \citenamefont {Libert}, \citenamefont {Litzinger},
  \citenamefont {Ljungvall}, \citenamefont {Louchart}, \citenamefont
  {Gottardo}, \citenamefont {Michelagnoli}, \citenamefont {M\"uller-Gatermann},
  \citenamefont {Napoli}, \citenamefont {Otsuka}, \citenamefont {Pillet},
  \citenamefont {Recchia}, \citenamefont {Rother}, \citenamefont {Sahin},
  \citenamefont {Siem}, \citenamefont {Sulignano}, \citenamefont {Togashi},
  \citenamefont {Tsunoda}, \citenamefont {Theisen},\ and\ \citenamefont
  {Valiente-Dobon}}]{Singh2018}%
  \BibitemOpen
  \bibfield  {author} {\bibinfo {author} {\bibfnamefont {P.}~\bibnamefont
  {Singh}},  \emph {et~al.},\ }\href {\doibase 10.1103/PhysRevLett.121.192501}
  {\bibfield  {journal} {\bibinfo  {journal} {Phys. Rev. Lett.}\ }\textbf
  {\bibinfo {volume} {121}},\ \bibinfo {pages} {192501} (\bibinfo {year}
  {2018})}\BibitemShut {NoStop}%
\bibitem [{\citenamefont {Hagen}\ \emph {et~al.}(2017)\citenamefont {Hagen},
  \citenamefont {G\"orgen}, \citenamefont {Korten}, \citenamefont {Grente},
  \citenamefont {Salsac}, \citenamefont {Farget}, \citenamefont {Ragnarsson},
  \citenamefont {Braunroth}, \citenamefont {Bruyneel}, \citenamefont
  {Celikovic}, \citenamefont {Cl\'ement}, \citenamefont {de~France},
  \citenamefont {Delaune}, \citenamefont {Dewald}, \citenamefont {Dijon},
  \citenamefont {Hackstein}, \citenamefont {Jacquot}, \citenamefont
  {Litzinger}, \citenamefont {Ljungvall}, \citenamefont {Louchart},
  \citenamefont {Michelagnoli}, \citenamefont {Napoli}, \citenamefont
  {Recchia}, \citenamefont {Rother}, \citenamefont {Sahin}, \citenamefont
  {Siem}, \citenamefont {Sulignano}, \citenamefont {Theisen},\ and\
  \citenamefont {Valiente-Dobon}}]{Hagen2017}%
  \BibitemOpen
  \bibfield  {author} {\bibinfo {author} {\bibfnamefont {T.~W.}\ \bibnamefont
  {Hagen}},  \emph {et~al.},\ }\href {\doibase 10.1103/PhysRevC.95.034302}
  {\bibfield  {journal} {\bibinfo  {journal} {Phys. Rev. C}\ }\textbf {\bibinfo
  {volume} {95}},\ \bibinfo {pages} {034302} (\bibinfo {year}
  {2017})}\BibitemShut {NoStop}%
\bibitem [{\citenamefont {Hagen}\ \emph {et~al.}(2018)\citenamefont {Hagen},
  \citenamefont {G{\"{o}}rgen}, \citenamefont {Korten}, \citenamefont {Grente},
  \citenamefont {Salsac}, \citenamefont {Farget}, \citenamefont {Braunroth},
  \citenamefont {Bruyneel}, \citenamefont {Celikovic}, \citenamefont
  {Cl{\'{e}}ment}, \citenamefont {de~France}, \citenamefont {Delaune},
  \citenamefont {Dewald}, \citenamefont {Dijon}, \citenamefont {Hackstein},
  \citenamefont {Jacquot}, \citenamefont {Litzinger}, \citenamefont
  {Ljungvall}, \citenamefont {Louchart}, \citenamefont {Michelagnoli},
  \citenamefont {Napoli}, \citenamefont {Recchia}, \citenamefont {Rother},
  \citenamefont {Sahin}, \citenamefont {Siem}, \citenamefont {Sulignano},
  \citenamefont {Theisen},\ and\ \citenamefont {Valiente-Dobon}}]{Hagen2018}%
  \BibitemOpen
  \bibfield  {author} {\bibinfo {author} {\bibfnamefont {T.~W.}\ \bibnamefont
  {Hagen}},  \emph {et~al.},\ }\href {\doibase 10.1140/epja/i2018-12482-9}
  {\bibfield  {journal} {\bibinfo  {journal} {Eur. Phys. J. A}\ }\textbf
  {\bibinfo {volume} {54}},\ \bibinfo {pages} {50} (\bibinfo {year}
  {2018})}\BibitemShut {NoStop}%
\bibitem [{\citenamefont {Akkoyun}(2012)\citenamefont {Akkoyun} \emph
  {et~al.}}]{AGATA}%
  \BibitemOpen
  \bibfield  {author} {\bibinfo {author} {\bibfnamefont {S.}~\bibnamefont
  {Akkoyun}} \emph {et~al.},\ }\href {\doibase 10.1016/j.nimA.2011.11.081}
  {\bibfield  {journal} {\bibinfo  {journal} {Nucl. Instrum. Methods Phys. Res.
  A}\ }\textbf {\bibinfo {volume} {668}},\ \bibinfo {pages} {26} (\bibinfo
  {year} {2012})}\BibitemShut {NoStop}%
\bibitem [{\citenamefont {Paschalis}\ \emph {et~al.}(2013)\citenamefont
  {Paschalis}, \citenamefont {Lee}, \citenamefont {Macchiavelli}, \citenamefont
  {Campbell}, \citenamefont {Cromaz}, \citenamefont {Gros}, \citenamefont
  {Pavan}, \citenamefont {Qian}, \citenamefont {Clark}, \citenamefont
  {Crawford}, \citenamefont {Doering}, \citenamefont {Fallon}, \citenamefont
  {Lionberger}, \citenamefont {Loew}, \citenamefont {Petri}, \citenamefont
  {Stezelberger}, \citenamefont {Zimmermann}, \citenamefont {Radford},
  \citenamefont {Lagergren}, \citenamefont {Weisshaar}, \citenamefont
  {Winkler}, \citenamefont {Glasmacher}, \citenamefont {Anderson},\ and\
  \citenamefont {Beausang}}]{Paschalis2013}%
  \BibitemOpen
  \bibfield  {author} {\bibinfo {author} {\bibfnamefont {S.}~\bibnamefont
  {Paschalis}},  \emph {et~al.},\ }\href {\doibase 10.1016/j.nimA.2013.01.009}
  {\bibfield  {journal} {\bibinfo  {journal} {Nucl. Instrum. Methods Phys. Res.
  A}\ }\textbf {\bibinfo {volume} {709}},\ \bibinfo {pages} {44} (\bibinfo
  {year} {2013})}\BibitemShut {NoStop}%
\bibitem [{\citenamefont {Stahl}\ \emph {et~al.}(2017)\citenamefont {Stahl},
  \citenamefont {Leske}, \citenamefont {Lettmann},\ and\ \citenamefont
  {Pietralla}}]{Stahl2017}%
  \BibitemOpen
  \bibfield  {author} {\bibinfo {author} {\bibfnamefont {C.}~\bibnamefont
  {Stahl}},  \emph {et~al.},\ }\href {\doibase
  https://doi.org/10.1016/j.cpc.2017.01.009} {\bibfield  {journal} {\bibinfo
  {journal} {Comp. Phys. Comm.}\ }\textbf {\bibinfo {volume} {214}},\ \bibinfo
  {pages} {174} (\bibinfo {year} {2017})}\BibitemShut {NoStop}%
\bibitem [{\citenamefont {Cl{\'{e}}ment}\ \emph {et~al.}(2017)\citenamefont
  {Cl{\'{e}}ment}, \citenamefont {Michelagnoli}, \citenamefont {de~France},
  \citenamefont {Li}, \citenamefont {Lemasson}, \citenamefont {{Barthe
  Dejean}}, \citenamefont {Beuzard}, \citenamefont {Bougault}, \citenamefont
  {Cacitti}, \citenamefont {Foucher}, \citenamefont {Fremont}, \citenamefont
  {Gangnant}, \citenamefont {Goupil}, \citenamefont {Houarner}, \citenamefont
  {Jean}, \citenamefont {Lefevre}, \citenamefont {Legeard}, \citenamefont
  {Legruel}, \citenamefont {Maugeais}, \citenamefont {M{\'{e}}nager},
  \citenamefont {M{\'{e}}nard}, \citenamefont {Munoz}, \citenamefont {Ozille},
  \citenamefont {Raine}, \citenamefont {Ropert}, \citenamefont {Saillant},
  \citenamefont {Spitaels}, \citenamefont {Tripon}, \citenamefont {Vallerand},
  \citenamefont {Voltolini}, \citenamefont {Korten}, \citenamefont {Salsac},
  \citenamefont {Theisen}, \citenamefont {Zieli{\'{n}}ska}, \citenamefont
  {Joannem}, \citenamefont {Karolak}, \citenamefont {Kebbiri}, \citenamefont
  {Lotode}, \citenamefont {Touzery}, \citenamefont {Walter}, \citenamefont
  {Korichi}, \citenamefont {Ljungvall}, \citenamefont {Lopez-Martens},
  \citenamefont {Ralet}, \citenamefont {Dosme}, \citenamefont {Grave},
  \citenamefont {Karkour}, \citenamefont {Lafay}, \citenamefont {Legay},
  \citenamefont {Kojouharov}, \citenamefont {Domingo-Pardo}, \citenamefont
  {Gadea}, \citenamefont {P{\'{e}}rez-Vidal}, \citenamefont {Civera},
  \citenamefont {Birkenbach}, \citenamefont {Eberth}, \citenamefont {Hess},
  \citenamefont {Lewandowski}, \citenamefont {Reiter}, \citenamefont {Nannini},
  \citenamefont {{De Angelis}}, \citenamefont {Jaworski}, \citenamefont {John},
  \citenamefont {Napoli}, \citenamefont {Valiente-Dob{\'{o}}n}, \citenamefont
  {Barrientos}, \citenamefont {Bortolato}, \citenamefont {Benzoni},
  \citenamefont {Bracco}, \citenamefont {Brambilla}, \citenamefont {Camera},
  \citenamefont {Crespi}, \citenamefont {Leoni}, \citenamefont {Million},
  \citenamefont {Pullia}, \citenamefont {Wieland}, \citenamefont {Bazzacco},
  \citenamefont {Lenzi}, \citenamefont {Lunardi}, \citenamefont {Menegazzo},
  \citenamefont {Mengoni}, \citenamefont {Recchia}, \citenamefont {Bellato},
  \citenamefont {Isocrate}, \citenamefont {{Egea Canet}}, \citenamefont
  {Didierjean}, \citenamefont {Duch{\^{e}}ne}, \citenamefont {Baumann},
  \citenamefont {Brucker}, \citenamefont {Dangelser}, \citenamefont {Filliger},
  \citenamefont {Friedmann}, \citenamefont {Gaudiot}, \citenamefont {Grapton},
  \citenamefont {Kocher}, \citenamefont {Mathieu}, \citenamefont {Sigward},
  \citenamefont {Thomas}, \citenamefont {Veeramootoo}, \citenamefont {Dudouet},
  \citenamefont {St{\'{e}}zowski}, \citenamefont {Aufranc}, \citenamefont
  {Aubert}, \citenamefont {Labiche}, \citenamefont {Simpson}, \citenamefont
  {Burrows}, \citenamefont {Coleman-Smith}, \citenamefont {Grant},
  \citenamefont {Lazarus}, \citenamefont {Morrall}, \citenamefont {Pucknell},
  \citenamefont {Boston}, \citenamefont {Judson}, \citenamefont
  {Lalovi{\'{c}}}, \citenamefont {Nyberg}, \citenamefont {Collado},
  \citenamefont {Gonz{\'{a}}lez}, \citenamefont {Kuti}, \citenamefont
  {Nyak{\'{o}}}, \citenamefont {Maj},\ and\ \citenamefont
  {Rudigier}}]{Clement2017}%
  \BibitemOpen
  \bibfield  {author} {\bibinfo {author} {\bibfnamefont {E.}~\bibnamefont
  {Cl{\'{e}}ment}},  \emph {et~al.},\ }\href {\doibase
  10.1016/j.nima.2017.02.063} {\bibfield  {journal} {\bibinfo  {journal} {Nucl.
  Instrum. Methods Phys. Res. A}\ }\textbf {\bibinfo {volume} {855}},\ \bibinfo
  {pages} {1} (\bibinfo {year} {2017})}\BibitemShut {NoStop}%
\bibitem [{\citenamefont {Vandebrouck}\ \emph {et~al.}(2016)\citenamefont
  {Vandebrouck}, \citenamefont {Lemasson}, \citenamefont {Rejmund},
  \citenamefont {Fremont}, \citenamefont {Pancin}, \citenamefont {Navin},
  \citenamefont {Michelagnoli}, \citenamefont {Goupil}, \citenamefont
  {Spitaels},\ and\ \citenamefont {Jacquot}}]{Vandebrouck2016}%
  \BibitemOpen
  \bibfield  {author} {\bibinfo {author} {\bibfnamefont {M.}~\bibnamefont
  {Vandebrouck}},  \emph {et~al.},\ }\href {\doibase
  10.1016/j.nima.2015.12.040} {\bibfield  {journal} {\bibinfo  {journal} {Nucl.
  Instrum. Methods Phys. Res. A}\ }\textbf {\bibinfo {volume} {812}},\ \bibinfo
  {pages} {112 } (\bibinfo {year} {2016})}\BibitemShut {NoStop}%
\bibitem [{\citenamefont {Ljungvall}\ \emph {et~al.}(2020)\citenamefont
  {Ljungvall}, \citenamefont {Pérez-Vidal}, \citenamefont {Lopez-Martens},
  \citenamefont {Michelagnoli}, \citenamefont {Clément}, \citenamefont
  {Dudouet}, \citenamefont {Gadea}, \citenamefont {Hess}, \citenamefont
  {Korichi}, \citenamefont {Labiche}, \citenamefont {Lalović}, \citenamefont
  {Li},\ and\ \citenamefont {Recchia}}]{LjungVall2020}%
  \BibitemOpen
  \bibfield  {author} {\bibinfo {author} {\bibfnamefont {J.}~\bibnamefont
  {Ljungvall}},  \emph {et~al.},\ }\href {\doibase
  https://doi.org/10.1016/j.nima.2019.163297} {\bibfield  {journal} {\bibinfo
  {journal} {Nucl. Instrum. Methods Phys. Res. A}\ }\textbf {\bibinfo {volume}
  {955}},\ \bibinfo {pages} {163297} (\bibinfo {year} {2020})}\BibitemShut
  {NoStop}%
\bibitem [{\citenamefont {Recchia}(2008)}]{PhDRecchia}%
  \BibitemOpen
  \bibfield  {author} {\bibinfo {author} {\bibfnamefont {F.}~\bibnamefont
  {Recchia}},\ }\emph {\bibinfo {title} {{In-beam test and imaging capabilities
  of the AGATA prototype detector}}},\ \href
  {https://npgroup.pd.infn.it/Tesi/PhD-thesisRecchia.pdf} {Ph.D. thesis},\
  \bibinfo  {school} {Universita Degli Studi Di Padova} (\bibinfo {year}
  {2008})\BibitemShut {NoStop}%
\bibitem [{\citenamefont {Recchia}\ \emph {et~al.}(2009)\citenamefont
  {Recchia}, \citenamefont {Bazzacco}, \citenamefont {Farnea}, \citenamefont
  {Gadea}, \citenamefont {Venturelli}, \citenamefont {Beck}, \citenamefont
  {Bednarczyk}, \citenamefont {Buerger}, \citenamefont {Dewald}, \citenamefont
  {Dimmock}, \citenamefont {Duchêne}, \citenamefont {Eberth}, \citenamefont
  {Faul}, \citenamefont {Gerl}, \citenamefont {Gernhaeuser}, \citenamefont
  {Hauschild}, \citenamefont {Holler}, \citenamefont {Jones}, \citenamefont
  {Korten}, \citenamefont {Kröll}, \citenamefont {Krücken}, \citenamefont
  {Kurz}, \citenamefont {Ljungvall}, \citenamefont {Lunardi}, \citenamefont
  {Maierbeck}, \citenamefont {Mengoni}, \citenamefont {Nyberg}, \citenamefont
  {Nelson}, \citenamefont {Pascovici}, \citenamefont {Reiter}, \citenamefont
  {Schaffner}, \citenamefont {Schlarb}, \citenamefont {Steinhardt},
  \citenamefont {Thelen}, \citenamefont {Ur}, \citenamefont {{Valiente
  Dobon}},\ and\ \citenamefont {Weißhaar}}]{Recchia2009}%
  \BibitemOpen
  \bibfield  {author} {\bibinfo {author} {\bibfnamefont {F.}~\bibnamefont
  {Recchia}},  \emph {et~al.},\ }\href {\doibase
  https://doi.org/10.1016/j.nima.2009.02.042} {\bibfield  {journal} {\bibinfo
  {journal} {Nucl. Instrum. Methods Phys. Res. A}\ }\textbf {\bibinfo {volume}
  {604}},\ \bibinfo {pages} {555} (\bibinfo {year} {2009})}\BibitemShut
  {NoStop}%
\bibitem [{\citenamefont {Söderström}\ \emph {et~al.}(2011)\citenamefont
  {Söderström}, \citenamefont {Recchia}, \citenamefont {Nyberg},
  \citenamefont {Al-Adili}, \citenamefont {Ataç}, \citenamefont {Aydin},
  \citenamefont {Bazzacco}, \citenamefont {Bednarczyk}, \citenamefont
  {Birkenbach}, \citenamefont {Bortolato}, \citenamefont {Boston},
  \citenamefont {Boston}, \citenamefont {Bruyneel}, \citenamefont {Bucurescu},
  \citenamefont {Calore}, \citenamefont {Colosimo}, \citenamefont {Crespi},
  \citenamefont {Dosme}, \citenamefont {Eberth}, \citenamefont {Farnea},
  \citenamefont {Filmer}, \citenamefont {Gadea}, \citenamefont {Gottardo},
  \citenamefont {Grave}, \citenamefont {Grebosz}, \citenamefont {Griffiths},
  \citenamefont {Gulmini}, \citenamefont {Habermann}, \citenamefont {Hess},
  \citenamefont {Jaworski}, \citenamefont {Jones}, \citenamefont {Joshi},
  \citenamefont {Judson}, \citenamefont {Kempley}, \citenamefont {Khaplanov},
  \citenamefont {Legay}, \citenamefont {Lersch}, \citenamefont {Ljungvall},
  \citenamefont {Lopez-Martens}, \citenamefont {Meczynski}, \citenamefont
  {Mengoni}, \citenamefont {Michelagnoli}, \citenamefont {Molini},
  \citenamefont {Napoli}, \citenamefont {Orlandi}, \citenamefont {Pascovici},
  \citenamefont {Pullia}, \citenamefont {Reiter}, \citenamefont {Sahin},
  \citenamefont {Smith}, \citenamefont {Strachan}, \citenamefont {Tonev},
  \citenamefont {Unsworth}, \citenamefont {Ur}, \citenamefont
  {Valiente-Dobón}, \citenamefont {Veyssiere},\ and\ \citenamefont
  {Wiens}}]{Soderstrom2011}%
  \BibitemOpen
  \bibfield  {author} {\bibinfo {author} {\bibfnamefont {P.-A.}\ \bibnamefont
  {Söderström}},  \emph {et~al.},\ }\href {\doibase
  https://doi.org/10.1016/j.nima.2011.02.089} {\bibfield  {journal} {\bibinfo
  {journal} {Nucl. Instrum. Methods Phys. Res. A}\ }\textbf {\bibinfo {volume}
  {638}},\ \bibinfo {pages} {96} (\bibinfo {year} {2011})}\BibitemShut
  {NoStop}%
\bibitem [{\citenamefont {Bhattacharyya}\ \emph {et~al.}(2008)\citenamefont
  {Bhattacharyya}, \citenamefont {Rejmund}, \citenamefont {Navin},
  \citenamefont {Caurier}, \citenamefont {Nowacki}, \citenamefont {Poves},
  \citenamefont {Chapman}, \citenamefont {O'Donnell}, \citenamefont {Gelin},
  \citenamefont {Hodsdon}, \citenamefont {Liang}, \citenamefont {Mittig},
  \citenamefont {Mukherjee}, \citenamefont {Rejmund}, \citenamefont {Rousseau},
  \citenamefont {Roussel-Chomaz}, \citenamefont {Spohr},\ and\ \citenamefont
  {Theisen}}]{Sam08}%
  \BibitemOpen
  \bibfield  {author} {\bibinfo {author} {\bibfnamefont {S.}~\bibnamefont
  {Bhattacharyya}},  \emph {et~al.},\ }\href {\doibase
  10.1103/PhysRevLett.101.032501} {\bibfield  {journal} {\bibinfo  {journal}
  {Phys. Rev. Lett.}\ }\textbf {\bibinfo {volume} {101}},\ \bibinfo {pages}
  {032501} (\bibinfo {year} {2008})}\BibitemShut {NoStop}%
\bibitem [{\citenamefont {Albers}\ \emph {et~al.}(2012)\citenamefont {Albers},
  \citenamefont {Warr}, \citenamefont {Nomura}, \citenamefont {Blazhev},
  \citenamefont {Jolie}, \citenamefont {M\"ucher}, \citenamefont {Bastin},
  \citenamefont {Bauer}, \citenamefont {Bernards}, \citenamefont {Bettermann},
  \citenamefont {Bildstein}, \citenamefont {Butterworth}, \citenamefont
  {Cappellazzo}, \citenamefont {Cederk\"all}, \citenamefont {Cline},
  \citenamefont {Darby}, \citenamefont {Das~Gupta}, \citenamefont {Daugas},
  \citenamefont {Davinson}, \citenamefont {De~Witte}, \citenamefont {Diriken},
  \citenamefont {Filipescu}, \citenamefont {Fiori}, \citenamefont {Fransen},
  \citenamefont {Gaffney}, \citenamefont {Georgiev}, \citenamefont
  {Gernh\"auser}, \citenamefont {Hackstein}, \citenamefont {Heinze},
  \citenamefont {Hess}, \citenamefont {Huyse}, \citenamefont {Jenkins},
  \citenamefont {Konki}, \citenamefont {Kowalczyk}, \citenamefont {Kr\"oll},
  \citenamefont {Kr\"ucken}, \citenamefont {Litzinger}, \citenamefont {Lutter},
  \citenamefont {Marginean}, \citenamefont {Mihai}, \citenamefont {Moschner},
  \citenamefont {Napiorkowski}, \citenamefont {Nara~Singh}, \citenamefont
  {Nowak}, \citenamefont {Otsuka}, \citenamefont {Pakarinen}, \citenamefont
  {Pfeiffer}, \citenamefont {Radeck}, \citenamefont {Reiter}, \citenamefont
  {Rigby}, \citenamefont {Robledo}, \citenamefont {Rodr\'{\i}guez-Guzm\'an},
  \citenamefont {Rudigier}, \citenamefont {Sarriguren}, \citenamefont {Scheck},
  \citenamefont {Seidlitz}, \citenamefont {Siebeck}, \citenamefont {Simpson},
  \citenamefont {Th\"ole}, \citenamefont {Thomas}, \citenamefont {Van~de
  Walle}, \citenamefont {Van~Duppen}, \citenamefont {Vermeulen}, \citenamefont
  {Voulot}, \citenamefont {Wadsworth}, \citenamefont {Wenander}, \citenamefont
  {Wimmer}, \citenamefont {Zell},\ and\ \citenamefont
  {Zielinska}}]{Albers2012}%
  \BibitemOpen
  \bibfield  {author} {\bibinfo {author} {\bibfnamefont {M.}~\bibnamefont
  {Albers}},  \emph {et~al.},\ }\href {\doibase 10.1103/PhysRevLett.108.062701}
  {\bibfield  {journal} {\bibinfo  {journal} {Phys. Rev. Lett.}\ }\textbf
  {\bibinfo {volume} {108}},\ \bibinfo {pages} {062701} (\bibinfo {year}
  {2012})}\BibitemShut {NoStop}%
\bibitem [{\citenamefont {Casten}(1990)}]{Casten1990}%
  \BibitemOpen
  \bibfield  {author} {\bibinfo {author} {\bibfnamefont {R.}~\bibnamefont
  {Casten}},\ }\href@noop {} {\emph {\bibinfo {title} {Nuclear Structure from a
  Simple Perspective}}}\ (\bibinfo  {publisher} {Oxford Science Publication},\
  \bibinfo {address} {New York},\ \bibinfo {year} {1990})\BibitemShut {NoStop}%
\bibitem [{\citenamefont {Flavigny}\ \emph {et~al.}(2017)\citenamefont
  {Flavigny}, \citenamefont {Doornenbal}, \citenamefont {Obertelli},
  \citenamefont {Delaroche}, \citenamefont {Girod}, \citenamefont {Libert},
  \citenamefont {Rodriguez}, \citenamefont {Authelet}, \citenamefont {Baba},
  \citenamefont {Calvet}, \citenamefont {Ch\^ateau}, \citenamefont {Chen},
  \citenamefont {Corsi}, \citenamefont {Delbart}, \citenamefont {Gheller},
  \citenamefont {Giganon}, \citenamefont {Gillibert}, \citenamefont {Lapoux},
  \citenamefont {Motobayashi}, \citenamefont {Niikura}, \citenamefont {Paul},
  \citenamefont {Rouss\'e}, \citenamefont {Sakurai}, \citenamefont
  {Santamaria}, \citenamefont {Steppenbeck}, \citenamefont {Taniuchi},
  \citenamefont {Uesaka}, \citenamefont {Ando}, \citenamefont {Arici},
  \citenamefont {Blazhev}, \citenamefont {Browne}, \citenamefont {Bruce},
  \citenamefont {Carroll}, \citenamefont {Chung}, \citenamefont {Cort\'es},
  \citenamefont {Dewald}, \citenamefont {Ding}, \citenamefont {Franchoo},
  \citenamefont {G\'orska}, \citenamefont {Gottardo}, \citenamefont
  {Jungclaus}, \citenamefont {Lee}, \citenamefont {Lettmann}, \citenamefont
  {Linh}, \citenamefont {Liu}, \citenamefont {Liu}, \citenamefont {Lizarazo},
  \citenamefont {Momiyama}, \citenamefont {Moschner}, \citenamefont {Nagamine},
  \citenamefont {Nakatsuka}, \citenamefont {Nita}, \citenamefont {Nobs},
  \citenamefont {Olivier}, \citenamefont {Orlandi}, \citenamefont {Patel},
  \citenamefont {Podoly\'ak}, \citenamefont {Rudigier}, \citenamefont {Saito},
  \citenamefont {Shand}, \citenamefont {S\"oderstr\"om}, \citenamefont
  {Stefan}, \citenamefont {Vaquero}, \citenamefont {Werner}, \citenamefont
  {Wimmer},\ and\ \citenamefont {Xu}}]{Flavigny2017}%
  \BibitemOpen
  \bibfield  {author} {\bibinfo {author} {\bibfnamefont {F.}~\bibnamefont
  {Flavigny}},  \emph {et~al.},\ }\href {\doibase
  10.1103/PhysRevLett.118.242501} {\bibfield  {journal} {\bibinfo  {journal}
  {Phys. Rev. Lett.}\ }\textbf {\bibinfo {volume} {118}},\ \bibinfo {pages}
  {242501} (\bibinfo {year} {2017})}\BibitemShut {NoStop}%
\bibitem [{\citenamefont {Gerst}\ \emph {et~al.}(2022)\citenamefont {Gerst},
  \citenamefont {Blazhev}, \citenamefont {Moschner}, \citenamefont
  {Doornenbal}, \citenamefont {Obertelli}, \citenamefont {Nomura},
  \citenamefont {Ebran}, \citenamefont {Hilaire}, \citenamefont {Libert},
  \citenamefont {Authelet}, \citenamefont {Baba}, \citenamefont {Calvet},
  \citenamefont {Ch\^ateau}, \citenamefont {Chen}, \citenamefont {Corsi},
  \citenamefont {Delbart}, \citenamefont {Gheller}, \citenamefont {Giganon},
  \citenamefont {Gillibert}, \citenamefont {Lapoux}, \citenamefont
  {Motobayashi}, \citenamefont {Niikura}, \citenamefont {Paul}, \citenamefont
  {Rouss\'e}, \citenamefont {Sakurai}, \citenamefont {Santamaria},
  \citenamefont {Steppenbeck}, \citenamefont {Taniuchi}, \citenamefont
  {Uesaka}, \citenamefont {Ando}, \citenamefont {Arici}, \citenamefont
  {Browne}, \citenamefont {Bruce}, \citenamefont {Caroll}, \citenamefont
  {Chung}, \citenamefont {Cort\'es}, \citenamefont {Dewald}, \citenamefont
  {Ding}, \citenamefont {Flavigny}, \citenamefont {Franchoo}, \citenamefont
  {G\'orska}, \citenamefont {Gottardo}, \citenamefont {Jolie}, \citenamefont
  {Jungclaus}, \citenamefont {Lee}, \citenamefont {Lettmann}, \citenamefont
  {Linh}, \citenamefont {Liu}, \citenamefont {Liu}, \citenamefont {Lizarazo},
  \citenamefont {Momiyama}, \citenamefont {Nagamine}, \citenamefont
  {Nakatsuka}, \citenamefont {Nita}, \citenamefont {Nobs}, \citenamefont
  {Olivier}, \citenamefont {Orlandi}, \citenamefont {Patel}, \citenamefont
  {Podoly\'ak}, \citenamefont {Rudigier}, \citenamefont {Saito}, \citenamefont
  {Shand}, \citenamefont {S\"oderstr\"om}, \citenamefont {Stefan},
  \citenamefont {Vaquero}, \citenamefont {Werner}, \citenamefont {Wimmer},\
  and\ \citenamefont {Xu}}]{Gerst2022}%
  \BibitemOpen
  \bibfield  {author} {\bibinfo {author} {\bibfnamefont {R.-B.}\ \bibnamefont
  {Gerst}},  \emph {et~al.},\ }\href {\doibase 10.1103/PhysRevC.105.024302}
  {\bibfield  {journal} {\bibinfo  {journal} {Phys. Rev. C}\ }\textbf {\bibinfo
  {volume} {105}},\ \bibinfo {pages} {024302} (\bibinfo {year}
  {2022})}\BibitemShut {NoStop}%
\bibitem [{\citenamefont {Nowacki}\ \emph {et~al.}(2016)\citenamefont
  {Nowacki}, \citenamefont {Poves}, \citenamefont {Caurier},\ and\
  \citenamefont {Bounthong}}]{Nowacki2016}%
  \BibitemOpen
  \bibfield  {author} {\bibinfo {author} {\bibfnamefont {F.}~\bibnamefont
  {Nowacki}},  \emph {et~al.},\ }\href {\doibase
  10.1103/PhysRevLett.117.272501} {\bibfield  {journal} {\bibinfo  {journal}
  {Phys. Rev. Lett.}\ }\textbf {\bibinfo {volume} {117}},\ \bibinfo {pages}
  {272501} (\bibinfo {year} {2016})}\BibitemShut {NoStop}%
\bibitem [{\citenamefont {Basu}(2020)\citenamefont {Basu}\ and\ \citenamefont
  {Cutchan}}]{Basu2020}%
  \BibitemOpen
  \bibfield  {author} {\bibinfo {author} {\bibfnamefont {S.}~\bibnamefont
  {Basu}}\ and\ \bibinfo {author} {\bibfnamefont {E.~M.}\ \bibnamefont
  {Cutchan}},\ }\href {\doibase https://doi.org/10.1016/j.nds.2020.04.001}
  {\bibfield  {journal} {\bibinfo  {journal} {Nucl. Data Sheets}\ }\textbf
  {\bibinfo {volume} {165}},\ \bibinfo {pages} {1} (\bibinfo {year}
  {2020})}\BibitemShut {NoStop}%
\bibitem [{\citenamefont {Elman}\ \emph {et~al.}(2017)\citenamefont {Elman},
  \citenamefont {Gade}, \citenamefont {Weisshaar}, \citenamefont {Barofsky},
  \citenamefont {Bazin}, \citenamefont {Bender}, \citenamefont {Bowry},
  \citenamefont {Hjorth-Jensen}, \citenamefont {Kemper}, \citenamefont
  {Lipschutz}, \citenamefont {Lunderberg}, \citenamefont {Sachmpazidi},
  \citenamefont {Terpstra}, \citenamefont {Walters}, \citenamefont
  {Westerberg}, \citenamefont {Williams},\ and\ \citenamefont
  {Wimmer}}]{Elman2017}%
  \BibitemOpen
  \bibfield  {author} {\bibinfo {author} {\bibfnamefont {B.}~\bibnamefont
  {Elman}},  \emph {et~al.},\ }\href {\doibase 10.1103/PhysRevC.96.044332}
  {\bibfield  {journal} {\bibinfo  {journal} {Phys. Rev. C}\ }\textbf {\bibinfo
  {volume} {96}},\ \bibinfo {pages} {044332} (\bibinfo {year}
  {2017})}\BibitemShut {NoStop}%
\bibitem [{\citenamefont {Sieja}\ \emph {et~al.}(2013)\citenamefont {Sieja},
  \citenamefont {Rodr\'{\i}guez}, \citenamefont {Kolos},\ and\ \citenamefont
  {Verney}}]{Sieja2013}%
  \BibitemOpen
  \bibfield  {author} {\bibinfo {author} {\bibfnamefont {K.}~\bibnamefont
  {Sieja}},  \emph {et~al.},\ }\href {\doibase 10.1103/PhysRevC.88.034327}
  {\bibfield  {journal} {\bibinfo  {journal} {Phys. Rev. C}\ }\textbf {\bibinfo
  {volume} {88}},\ \bibinfo {pages} {034327} (\bibinfo {year}
  {2013})}\BibitemShut {NoStop}%
\bibitem [{\citenamefont {Litzinger}\ \emph {et~al.}(2015)\citenamefont
  {Litzinger}, \citenamefont {Blazhev}, \citenamefont {Dewald}, \citenamefont
  {Didierjean}, \citenamefont {Duch\^ene}, \citenamefont {Fransen},
  \citenamefont {Lozeva}, \citenamefont {Sieja}, \citenamefont {Verney},
  \citenamefont {de~Angelis}, \citenamefont {Bazzacco}, \citenamefont
  {Birkenbach}, \citenamefont {Bottoni}, \citenamefont {Bracco}, \citenamefont
  {Braunroth}, \citenamefont {Cederwall}, \citenamefont {Corradi},
  \citenamefont {Crespi}, \citenamefont {D\'esesquelles}, \citenamefont
  {Eberth}, \citenamefont {Ellinger}, \citenamefont {Farnea}, \citenamefont
  {Fioretto}, \citenamefont {Gernh\"auser}, \citenamefont {Goasduff},
  \citenamefont {G\"orgen}, \citenamefont {Gottardo}, \citenamefont {Grebosz},
  \citenamefont {Hackstein}, \citenamefont {Hess}, \citenamefont {Ibrahim},
  \citenamefont {Jolie}, \citenamefont {Jungclaus}, \citenamefont {Kolos},
  \citenamefont {Korten}, \citenamefont {Leoni}, \citenamefont {Lunardi},
  \citenamefont {Maj}, \citenamefont {Menegazzo}, \citenamefont {Mengoni},
  \citenamefont {Michelagnoli}, \citenamefont {Mijatovic}, \citenamefont
  {Million}, \citenamefont {M\"oller}, \citenamefont {Modamio}, \citenamefont
  {Montagnoli}, \citenamefont {Montanari}, \citenamefont {Morales},
  \citenamefont {Napoli}, \citenamefont {Niikura}, \citenamefont {Pollarolo},
  \citenamefont {Pullia}, \citenamefont {Quintana}, \citenamefont {Recchia},
  \citenamefont {Reiter}, \citenamefont {Rosso}, \citenamefont {Sahin},
  \citenamefont {Salsac}, \citenamefont {Scarlassara}, \citenamefont
  {S\"oderstr\"om}, \citenamefont {Stefanini}, \citenamefont {Stezowski},
  \citenamefont {Szilner}, \citenamefont {Theisen}, \citenamefont
  {Valiente~Dob\'on}, \citenamefont {Vandone},\ and\ \citenamefont
  {Vogt}}]{Litzinger2015}%
  \BibitemOpen
  \bibfield  {author} {\bibinfo {author} {\bibfnamefont {J.}~\bibnamefont
  {Litzinger}},  \emph {et~al.},\ }\href {\doibase 10.1103/PhysRevC.92.064322}
  {\bibfield  {journal} {\bibinfo  {journal} {Phys. Rev. C}\ }\textbf {\bibinfo
  {volume} {92}},\ \bibinfo {pages} {064322} (\bibinfo {year}
  {2015})}\BibitemShut {NoStop}%
\bibitem [{\citenamefont {Wolf}(1976)\citenamefont {Wolf}\ and\ \citenamefont
  {Cheifetz}}]{Wolf1976}%
  \BibitemOpen
  \bibfield  {author} {\bibinfo {author} {\bibfnamefont {A.}~\bibnamefont
  {Wolf}}\ and\ \bibinfo {author} {\bibfnamefont {E.}~\bibnamefont
  {Cheifetz}},\ }\href {\doibase 10.1103/PhysRevLett.36.1072} {\bibfield
  {journal} {\bibinfo  {journal} {Phys. Rev. Lett.}\ }\textbf {\bibinfo
  {volume} {36}},\ \bibinfo {pages} {1072} (\bibinfo {year}
  {1976})}\BibitemShut {NoStop}%
\bibitem [{\citenamefont {Fogelberg}\ \emph {et~al.}(1986)\citenamefont
  {Fogelberg}, \citenamefont {Stone}, \citenamefont {Gill}, \citenamefont
  {Mach}, \citenamefont {Warner}, \citenamefont {Aprahamian},\ and\
  \citenamefont {Rehfield}}]{Fogelberg1986}%
  \BibitemOpen
  \bibfield  {author} {\bibinfo {author} {\bibfnamefont {B.}~\bibnamefont
  {Fogelberg}},  \emph {et~al.},\ }\href {\doibase
  https://doi.org/10.1016/0375-9474(86)90244-7} {\bibfield  {journal} {\bibinfo
   {journal} {Nucl. Phys. A}\ }\textbf {\bibinfo {volume} {451}},\ \bibinfo
  {pages} {104} (\bibinfo {year} {1986})}\BibitemShut {NoStop}%
\bibitem [{\citenamefont {Astier}\ \emph {et~al.}(2012)\citenamefont {Astier},
  \citenamefont {Porquet}, \citenamefont {Theisen}, \citenamefont {Verney},
  \citenamefont {Deloncle}, \citenamefont {Houry}, \citenamefont {Lucas},
  \citenamefont {Azaiez}, \citenamefont {Barreau}, \citenamefont {Curien},
  \citenamefont {Dorvaux}, \citenamefont {Duch\^ene}, \citenamefont {Gall},
  \citenamefont {Redon}, \citenamefont {Rousseau},\ and\ \citenamefont
  {St\'ezowski}}]{Astier2012}%
  \BibitemOpen
  \bibfield  {author} {\bibinfo {author} {\bibfnamefont {A.}~\bibnamefont
  {Astier}},  \emph {et~al.},\ }\href {\doibase 10.1103/PhysRevC.85.054316}
  {\bibfield  {journal} {\bibinfo  {journal} {Phys. Rev. C}\ }\textbf {\bibinfo
  {volume} {85}},\ \bibinfo {pages} {054316} (\bibinfo {year}
  {2012})}\BibitemShut {NoStop}%
\bibitem [{\citenamefont {McDonald}(1973)\citenamefont {McDonald}\ and\
  \citenamefont {Kerek}}]{McDonald1973}%
  \BibitemOpen
  \bibfield  {author} {\bibinfo {author} {\bibfnamefont {J.}~\bibnamefont
  {McDonald}}\ and\ \bibinfo {author} {\bibfnamefont {A.}~\bibnamefont
  {Kerek}},\ }\href {\doibase https://doi.org/10.1016/0375-9474(73)90544-7}
  {\bibfield  {journal} {\bibinfo  {journal} {Nucl. Phys. A}\ }\textbf
  {\bibinfo {volume} {206}},\ \bibinfo {pages} {417} (\bibinfo {year}
  {1973})}\BibitemShut {NoStop}%
\bibitem [{\citenamefont {Hughes}\ \emph {et~al.}(2005)\citenamefont {Hughes},
  \citenamefont {Zamfir}, \citenamefont {Radford}, \citenamefont {Gross},
  \citenamefont {Barton}, \citenamefont {Baktash}, \citenamefont {Caprio},
  \citenamefont {Casten}, \citenamefont {Galindo-Uribarri}, \citenamefont
  {Hausladen}, \citenamefont {McCutchan}, \citenamefont {Ressler},
  \citenamefont {Shapira}, \citenamefont {Stracener},\ and\ \citenamefont
  {Yu}}]{Hughes2005}%
  \BibitemOpen
  \bibfield  {author} {\bibinfo {author} {\bibfnamefont {R.~O.}\ \bibnamefont
  {Hughes}},  \emph {et~al.},\ }\href {\doibase 10.1103/PhysRevC.71.044311}
  {\bibfield  {journal} {\bibinfo  {journal} {Phys. Rev. C}\ }\textbf {\bibinfo
  {volume} {71}},\ \bibinfo {pages} {044311} (\bibinfo {year}
  {2005})}\BibitemShut {NoStop}%
\bibitem [{\citenamefont {Roberts}\ \emph {et~al.}(2013)\citenamefont
  {Roberts}, \citenamefont {Bruce}, \citenamefont {Browne}, \citenamefont
  {Mărginean}, \citenamefont {Alexander}, \citenamefont {Alharbi},
  \citenamefont {Bucurescu}, \citenamefont {Deleanu}, \citenamefont {Delion},
  \citenamefont {Filipescu}, \citenamefont {Fraile}, \citenamefont {Gheorghe},
  \citenamefont {Ghiţă}, \citenamefont {Glodariu}, \citenamefont {Ivanova},
  \citenamefont {Kisyov}, \citenamefont {Mărginean}, \citenamefont {Mason},
  \citenamefont {Mihai}, \citenamefont {Mulholland}, \citenamefont {Negret},
  \citenamefont {Niţă}, \citenamefont {Olaizola}, \citenamefont {Pascu},
  \citenamefont {Söderström}, \citenamefont {Regan}, \citenamefont {Sava},
  \citenamefont {Stroe}, \citenamefont {Toma},\ and\ \citenamefont
  {Townsley}}]{Roberts2013}%
  \BibitemOpen
  \bibfield  {author} {\bibinfo {author} {\bibfnamefont {O.}~\bibnamefont
  {Roberts}},  \emph {et~al.},\ }\href {\doibase
  https://doi.org/10.5506/APhysPolB.44.403} {\bibfield  {journal} {\bibinfo
  {journal} {Acta Phys. Pol. B}\ }\textbf {\bibinfo {volume} {44}},\ \bibinfo
  {pages} {403} (\bibinfo {year} {2013})}\BibitemShut {NoStop}%
\bibitem [{\citenamefont {Genevey}\ \emph {et~al.}(2001)\citenamefont
  {Genevey}, \citenamefont {Pinston}, \citenamefont {Foin}, \citenamefont
  {Rejmund}, \citenamefont {Casten}, \citenamefont {Faust},\ and\ \citenamefont
  {Oberstedt}}]{Genevey2001}%
  \BibitemOpen
  \bibfield  {author} {\bibinfo {author} {\bibfnamefont {J.}~\bibnamefont
  {Genevey}},  \emph {et~al.},\ }\href {\doibase 10.1103/PhysRevC.63.054315}
  {\bibfield  {journal} {\bibinfo  {journal} {Phys. Rev. C}\ }\textbf {\bibinfo
  {volume} {63}},\ \bibinfo {pages} {054315} (\bibinfo {year}
  {2001})}\BibitemShut {NoStop}%
\bibitem [{\citenamefont {Biswas}\ \emph {et~al.}(2016)\citenamefont {Biswas},
  \citenamefont {Palit}, \citenamefont {Navin}, \citenamefont {Rejmund},
  \citenamefont {Bisoi}, \citenamefont {Sarkar}, \citenamefont {Sarkar},
  \citenamefont {Bhattacharyya}, \citenamefont {Biswas}, \citenamefont
  {Caama{\~{n}}o}, \citenamefont {Carpenter}, \citenamefont {Choudhury},
  \citenamefont {Cl{\'{e}}ment}, \citenamefont {Danu}, \citenamefont {Delaune},
  \citenamefont {Farget}, \citenamefont {de~France}, \citenamefont {Hota},
  \citenamefont {Jacquot}, \citenamefont {Lemasson}, \citenamefont
  {Mukhopadhyay}, \citenamefont {Nanal}, \citenamefont {Pillay}, \citenamefont
  {Saha}, \citenamefont {Sethi}, \citenamefont {Singh}, \citenamefont
  {Srivastava},\ and\ \citenamefont {Tandel}}]{Biswas2016}%
  \BibitemOpen
  \bibfield  {author} {\bibinfo {author} {\bibfnamefont {S.}~\bibnamefont
  {Biswas}},  \emph {et~al.},\ }\href {\doibase 10.1103/PhysRevC.93.034324}
  {\bibfield  {journal} {\bibinfo  {journal} {Phys. Rev. C}\ }\textbf {\bibinfo
  {volume} {93}},\ \bibinfo {pages} {034324} (\bibinfo {year}
  {2016})}\BibitemShut {NoStop}%
\end{thebibliography}%

\end{document}